\documentclass[twocolumn, trackchanges]{aastex63}

\usepackage[T1]{fontenc}
\usepackage{ae,aecompl}
\usepackage{graphicx}	% Including figure files
\usepackage{amsmath}	% Advanced maths commands
\usepackage{amssymb}	% Extra maths symbols
\usepackage{longtable}
\usepackage[bottom]{footmisc}
\usepackage{comment}
\usepackage{bm}
\newcommand{\co}{C$^{18}$O}
\newcommand{\tbol}{$T_{\text{bol}}$}
\newcommand{\lbol}{$L_{\text{bol}}$}
\newcommand{\radeff}{$R_{\text{eff}}$}
\newcommand{\skipthis}[1]{}

\shorttitle{Evolution \& Kinematics of \co\ Envelopes}
\shortauthors{Heimsoth et al.}
\begin{document}

\title{Evolution and Kinematics of Protostellar Envelopes in the Perseus Molecular Cloud}

\author{Daniel J. Heimsoth}
\affiliation{Department of Astronomy, Yale University, New Haven, CT, USA}
\affiliation{Center for Astrophysics | Harvard \& Smithsonian, 60 Garden Street, Cambridge, MA, USA}

\author{Ian W. Stephens}
\affiliation{Department of Earth, Environment, and Physics, Worcester State University, Worcester, MA 01602, USA}
\affiliation{Center for Astrophysics | Harvard \& Smithsonian, 60 Garden Street, Cambridge, MA, USA}

\author{H\'ector G. Arce}
\affiliation{Department of Astronomy, Yale University, New Haven, CT, USA}

\author{Tyler L. Bourke}
\affiliation{SKA Observatory, Jodrell Bank, Lower Withington, Macclesfield, Cheshire SK11 9FT, UK}
\affiliation{Center for Astrophysics | Harvard \& Smithsonian, 60 Garden Street, Cambridge, MA, USA}

\author{Philip C. Myers}
\affiliation{Center for Astrophysics | Harvard \& Smithsonian, 60 Garden Street, Cambridge, MA, USA}

\author{Michael M. Dunham}
\affiliation{Department of Physics, State University of New York at Fredonia, 280 Central Avenue, Fredonia, NY 14063, USA}



\begin{abstract}

We present a comprehensive analysis on the evolution of envelopes surrounding protostellar systems in the Perseus molecular cloud using data from the MASSES survey. We focus our attention to the \co(2--1) spectral line, and we characterize the shape, size, and orientation of 54 envelopes and measure their fluxes, velocity gradients, and line widths. To look for evolutionary trends, we compare these parameters to the bolometric temperature \tbol, a tracer of protostellar age. We find evidence that the angular difference between the elongation angle of the \co\ envelope and the outflow axis direction generally becomes increasingly perpendicular with increasing \tbol, suggesting the envelope evolution is directly affected by the outflow evolution. We show that this angular difference changes at \tbol\ = $53 \pm 20$\,K, which includes the conventional delineation between Class~0 and~I protostars of 70\,K. We compare the \co\ envelopes with larger gaseous structures in other molecular clouds and show that the velocity gradient increases with decreasing radius ($|\mathcal{G}| \sim R^{-0.72 \pm 0.06}$). From the velocity gradients we show that the specific angular momentum follows a power law fit $J/M \propto R^{1.83 \pm 0.05}$ for scales from 1pc down to $\sim$500 au, and we cannot rule out a possible flattening out at radii smaller than $\sim$1000 au.

\end{abstract}



\section{Introduction} \label{sec:intro}

Stars form in dense molecular clouds through the gravitational collapse of molecular gas. The process of mass accretion during the earliest stages of protostar development sets the final characteristics of the resultant star and thus is a crucial mechanism to study. Protostars have been categorized into different classes based on observational signatures of evolutionary stage \citep[e.g.,][]{Andre1993}. Class 0 protostars are the youngest, rapidly accreting mass from a dense, massive envelope. Class I protostars are in the later accretion phase, slowly gaining the rest of the mass from the envelope as it dissipates. By the time a protostar becomes Class II, all that is left is the accretion disk surrounding the star, with the envelope completely dissolved. Protostars can be characterized by the different gaseous structures that they are embedded in, namely the disk, envelope, and core. Disks, on scales of roughly $\sim$100 au, feed the accreting protostar and are the precursors to planetary systems. For Class 0 and Class I protostars, the disk is contained within a larger envelope of gas and dust, on the scale of $\sim$10$^2$-10$^4$ au.

The protostellar envelope plays an important role in feeding gas infalling from the surrounding molecular cloud core onto the accretion disk. Thus, the properties and evolution of the envelope are integral to the mass accretion process and to the eventual final mass of the star.  At the same time, protostars are prevented from accreting all the infalling gas due to the presence of bipolar outflows, which carry away excess angular momentum, allowing a fraction of the remaining infalling gas to accrete onto the protostar. The envelope is likely an important scale to probe angular momentum transfer, especially since angular momentum may be constant for scales smaller than $\sim$5000 au \citep[e.g.,][]{Belloche2013}.

\begin{figure}
     \centering
     \includegraphics[width=\linewidth]{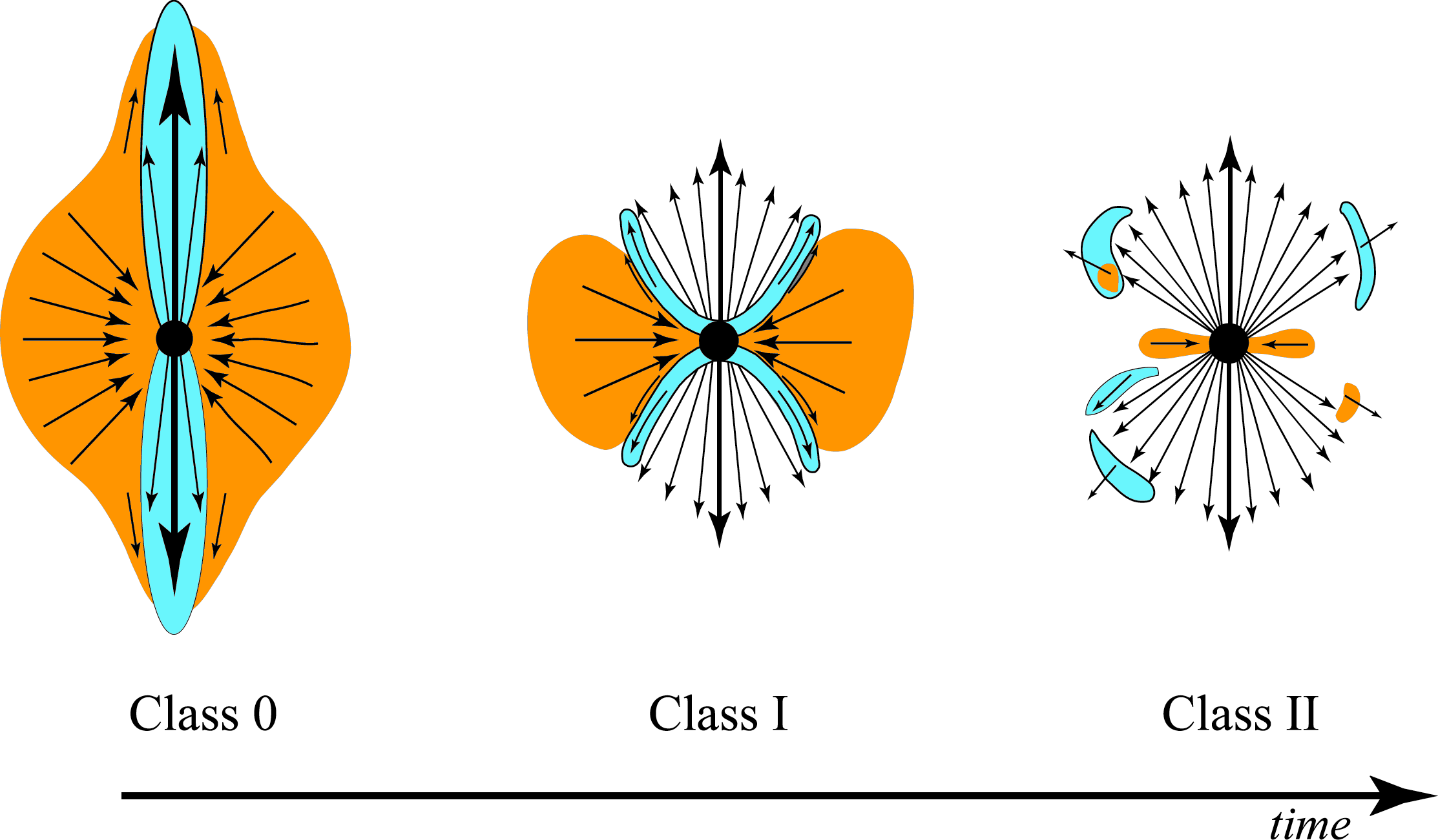}
     \caption{Cartoon adapted from Arce \& Sargeant (2006). During the Class 0 stage, some of the protostellar envelope (orange) is swept up with the outflow (cyan). At the Class I stage, the outflow has cleared out a cavity and the envelope only appears perpendicular to the outflow. By the Class II stage (not probed in this paper), the envelope and outflow have mostly dissipated.
     }
     \label{fig:arcecartoon}
 \end{figure}

The study of \cite{Arce_2006} was one of the first high-resolution spectral line and continuum surveys to investigate how envelopes and their outflows evolve through time. They looked at nine low-mass protostars at different evolutionary stages from Class 0 to Class II within 500 pc of the Sun. They concluded that the bipolar outflows have a major physical and chemical impact on the star formation process. They also proposed a mechanism where at young ages (Class 0) the dense gas in the envelope is entrained by the outflows. As the protostar ages, the outflows widen, concentrating the dense gas in structures perpendicular to the outflows. A cartoon of their prediction is shown in Figure~\ref{fig:arcecartoon}.

Because the analysis in \cite{Arce_2006} was only on nine sources, the evidence for their conclusions is marginal. Further, they chose bright sources from many different clouds, introducing an inherent selection bias. More recently, the Mass Assembly of Stellar Systems and Their Evolution with the SMA (MASSES) survey imaged the known 74 Class 0/I protostars in the Perseus molecular cloud (henceforth, Perseus). This survey used the Submillimeter Array (SMA) interferometer at 1.3\,mm and 870\,$\mu$m, mapping the continuum and spectral lines (including CO, $^{13}$CO, \co, and N$_2$D$^+$) at resolutions down to $\sim$1$\arcsec$ \citep{Stephens_2018,Stephens2019}. \cite{Zucker_2018} measured the distance to Perseus to be $\sim$300 pc using a combination of spectral line data, stellar photometry, and astrometric data.   The MASSES survey chose the Perseus molecular cloud because of its close proximity to Earth and its large amount of protostars to statistically constrain star formation processes \citep{Stephens_2018}. As a continuation of this work, we pose the following question: How do the bipolar outflows, detected very clearly in the CO line, affect the envelope, and hence the accretion of gas onto the protostar? As the MASSES sample of protostars is substantially larger than the sample of \cite{Arce_2006}, the following analysis minimizes selection and small sample size biases. 

Moreover, MASSES data also provides information about the kinematics of each individual envelope. The relation of how velocity gradient and specific angular momenta change with size scale has been analyzed quite a bit in the past \citep[e.g.,][]{Goodman_1993}.  We therefore use MASSES kinematic data to estimate velocity gradients and specific angular momenta of the envelopes and put them in context of larger scale star-forming regions from other studies.

In this paper, we analyze the MASSES survey data to characterize the \co(2--1) (henceforth, just \co) envelopes of protostars in the Perseus cloud, compare them to their associated outflows, and analyze the velocity gradients and line widths. We look for evolutionary trends and compare to measurements at larger scales. We define the sources and itemize the methods used in the data analysis in Section \ref{sec:data_methods}. In \S \ref{sec:analysis} we analyze the envelope shapes, intensities and orientations and conduct a multi-scale kinematics analysis using protostellar systems from other studies, and in \S \ref{sec:discussion} we give a discussion. Finally, in \S \ref{sec:conclusion}  we summarize our results.

\section{Data and Methods} \label{sec:data_methods}

\subsection{Observations \& Bolometric Temperature} \label{sec:sources}

\begin{figure}
    \centering
    \includegraphics[width=\linewidth]{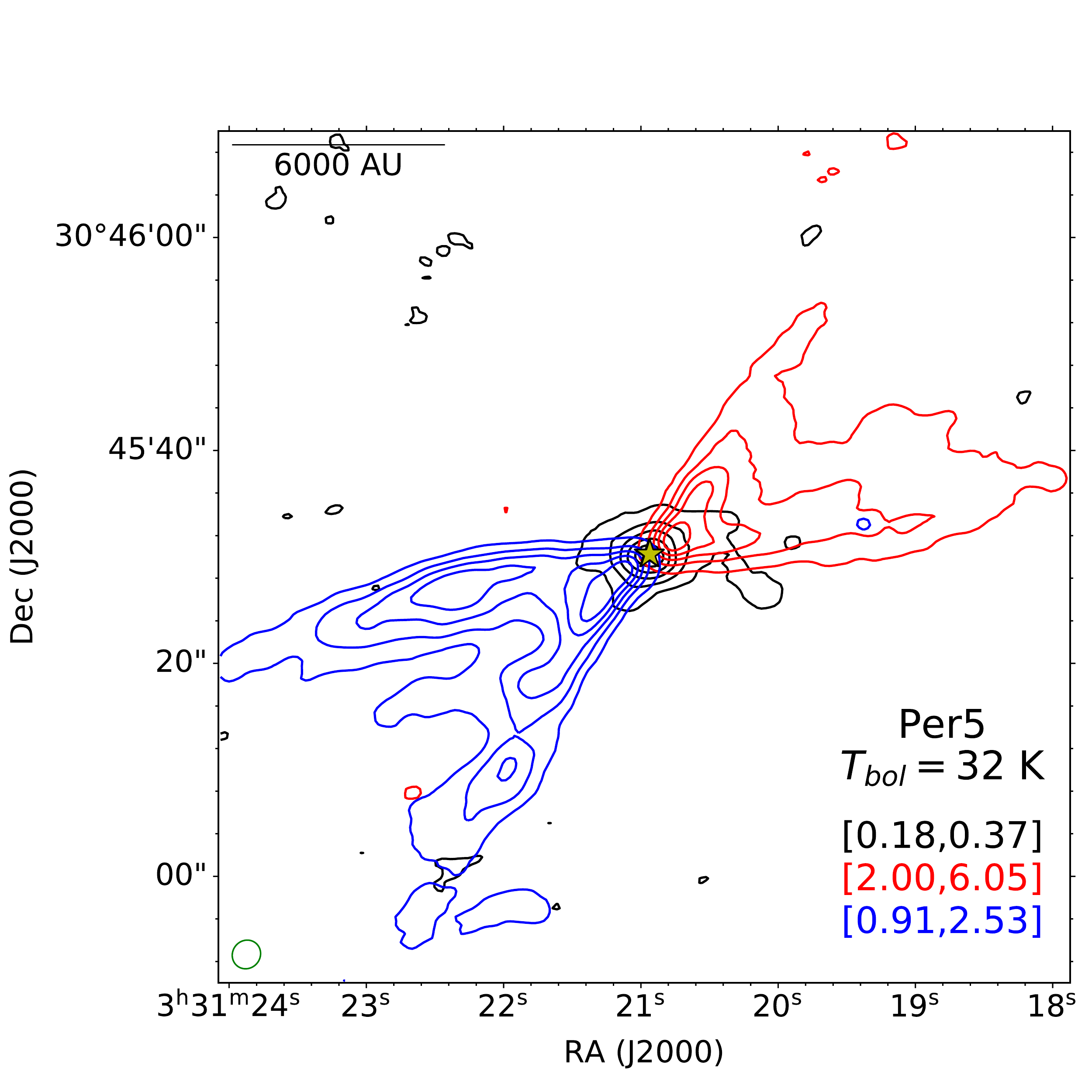}
    \caption{Per-emb-5 contour map, with \co\ (black), red-shifted CO (red), and blue-shifted CO (blue) contours. It is clear that the \co\ envelope is elongated along the direction of the CO outflow. The brackets are the first contour value and subsequent step in Jy km/s for the \co\ (red CO, blue CO) contours. The yellow star is the VLA continuum source detected by the VANDAM survey \citep{Tobin_2016}. The green ellipse in the lower left corner of the map is the synthesized beam of the SMA observations.}
    \label{fig:per5_img}
\end{figure}

All protostars in this study are in the Perseus molecular cloud, which has the largest count of Class 0/I protostars of all clouds within 350\,pc of Earth \citep{Dunham2015}. For this analysis, we take the nominal distance of Perseus from the Earth to be 300\,pc for all of our calculations \citep{Zucker_2018}. We use observations from the full MASSES survey, which combines the SMA's subcompact and extended array configuration \citep{Stephens2019}. The observations predominately used in this study are of the \co(2-1) spectral line, which traces the protostellar envelopes \citep{Frimann2017}. The \co\ envelopes were studied alongside CO(2--1) (henceforth CO) outflows, which were also observed using the SMA through the MASSES survey \citep{Stephens_2017}. Many protostars in this study are part of multiple systems, as detected and cataloged by the VLA Nascent Disk and Multiplicity (VANDAM) Perseus survey down to a resolution of $\sim$19 au \citep{Tobin_2016}. Data on source bolometric temperature and luminosity (\tbol\ and \lbol) and their associated errors were taken from the VANDAM survey \citep{Tobin_2016}. Figure \ref{fig:per5_img} shows a typical Class 0 object, Per-emb-5, in the \co\ and CO lines.

Table \ref{table:table1} gives the identifying characteristics of the 54 protostars used in this study, collected from \citep{Tobin_2016} and \citep{Stephens_2018}. Of these, 26 are Class 0, 11 are Class I, and seven are at the boundary of these two classifications (denoted as Class 0/I in Table \ref{table:table1} and the rest of this paper). The following analysis does not explicitly use class as a data point of interest or comparison, instead choosing to use \tbol\ for the ``age" of the system (see the following paragraphs for a discussion of \tbol). Class 0 and Class I protostars were deliberately chosen because more evolved Class II/III sources (i.e., T Tauri stars) have little or no envelope detectable in the \co\ spectral line. 

\tbol\ is used as a proxy for the age of each protostar \citep{1993ApJ...413L..47M}. \tbol\ is defined to be the temperature of a blackbody which has the same flux-weighted mean frequency as the spectral energy distribution (SED) of the source. While we report the errors in \tbol\ from \citet{Tobin_2016}, we do not show or use these errors in our plots. These errors only propagate flux uncertainties, while the dominant error is expected to be due to the incomplete SED coverage \citep{Enoch2009}. These errors only suggest whether the \tbol\ measurements are significant purely based on flux measurements, but the quoted uncertainties have no bearing on whether we trust the \tbol\ measured toward one protostar more than the other. When considering incomplete SED coverage, the typical error in \tbol\ is expected to be about 20\% \citep{Enoch2009}.

The age separation between Class~0 and~I protostars is typically taken to be at \tbol\ = 70~K \citep{Chen1995}. However, the observed \tbol\ can change considerably depending on the inclination of the source. \citet{Tobin_2016} gave the approximate classification of each protostar, which we report in Table~\ref{table:table1}. The lifetime of each protostellar class is highly uncertain. The best estimates based on population synthesis is about 0.15 and 0.3\,Myr for Class~0 and~I protostars, respectively \citep{Dunham2015}. However, these ages could be considerably shorter if populations are not in steady state \citep[$\sim$0.05 and 0.09\,Myr, respectively;][]{Kristensen2018}.

%Even within the sample studied in this paper, we start to see the dissolution of the \co\ envelopes for sources with higher \tbol\ (see Section \ref{sec:analysis}).

\subsection{Data Cubes \& Methods} \label{sec:methods}

As mentioned, the \co(2-1) data comes from the full MASSES data survey \citep{Stephens2019}. The data products are position-position-velocity cubes in FITS format. We specifically used the data cubes mapped using the Briggs' robust parameter of 1. The pixel size for all maps is 0.4$\arcsec$,  and the velocity channels have a width of 0.2\,km/s. The angular resolution (measured as the geometric mean of the major and minor axes of the synthesized beam) ranged from observation to observation \citep{Stephens2019}, with resolutions ranging from 1.4$\arcsec$ to 4.0$\arcsec$. The median resolution was 2.6$\arcsec$, corresponding to $\sim$780 au,

We made moment 0 (integrated intensity) maps of each source using the \texttt{moment()} method from the Python package \texttt{SpectralCube}\footnote{See \href{https://spectral-cube.readthedocs.io/en/latest/index.html}{spectral-cube.readthedocs.io}}. This function calculates the moment identically to the \texttt{immoments} method in Common Astronomy Software Applications package \citep[CASA;][]{McMullin2007}, integrating over velocity channels for each pixel. We then calculated the velocity and line width maps by fitting a 1D Gaussian to each pixel's spectrum along the velocity axis. We only kept pixels with less than 50\% line width errors. To extract useful data from the velocity and line width maps, we needed to isolate the \co\ envelope, so we masked the emission using the \texttt{astropy.stats.sigma\_clip} method with a 3-$\sigma$ cut above the RMS noise in the moment 0 map. The moment 0, velocity, and line width maps for the protostars studied in this paper are shown in Appendix~\ref{appendix:A}.

To measure the shape and flux of the \co\ envelopes, we fit a 2-dimensional Gaussian model onto each unmasked moment 0 map using CASA. We used the fitting functionality of the CASA \texttt{viewer()} on a polygon region around the \co\ envelope. From this, we obtained the position angle of the elongated axis of the envelope and the full-width half maximum (FWHM) major and minor axes of the Gaussian fit. We only consider fits in which the protostar is located within the FWHM of the \co\ envelope. Some position angles have errors higher than 45$^\circ$ (typically circular envelopes); while we include these fits in our tables and figures, they are not used in any of our statistical analyses. It may be noted that some envelopes have an X-like morphology (see for e.g., Per-emb-2 in Fig. \ref{fig:gallery_mom0}) which could affect the fitting. The Gaussian fits, however, are only sensitive to the central emission, and thus the outer emission shape has little to no effect on the fit. This can be best shown by again referring to Fig. \ref{fig:gallery_mom0}.

Before calculating the total integrated flux of each source, we removed small regions of emission that were not associated with the protostar's envelope. We did this by using the \texttt{remove\_small\_holes()} method in the Python package \texttt{skimage.morphology}\footnote{\href{https://scikit-image.org}{scikit-image.org}}. The combination of the 3-$\sigma$ mask and hole removal we henceforth call the ``clean mask." For the remainder of this study, all analysis was done using the ``clean" 3-$\sigma$ masked versions of the moment 0, velocity, and line width maps, unless explicitly stated otherwise.

To calculate total integrated fluxes, we used the clean mask on the moment 0 maps and calculated the total integrated flux by summing the flux of each pixel. The peak integrated flux for each source was also taken from these clean mask moment 0 maps by choosing the pixel with the maximum pixel flux value. The errors on the peak integrated flux are taken to be the standard deviation of all background pixels (i.e., pixels that are masked in the 3-$\sigma$ maps), while the errors of the total integrated flux were calculated as the peak integrated flux error multiplied by the square root of the number of beams per envelope area.

From the total integrated flux maps, in principle it is straightforward to calculate the local thermodynamic equilibrium gas mass of the envelope by assuming an excitation temperature, an abundance ratio, and optically thin emission (i.e., a local thermodynamic equilibrium mass). However, the abundance of carbon monoxide in the gas phase in dense and cold circumstellar envelopes is poorly constrained, as it differs significantly from the standard values used in the interstellar medium due to freeze-out onto dust grains \citep[e.g.,][]{Yildiz2013}. Indeed, there may not be an accurate canonical abundance ratio for envelopes, as episodic accretion may be ubiquitous and will change the fraction of carbon monoxide gas that will be depleted \citep[e.g.,][]{Frimann2017}. Moreover, \co\ may be optically thick, and we do not have an accurate way to estimate the \co\ optical depth with MASSES data. Because of the considerable uncertainties in both the abundances and optical depths, we do not estimate envelope gas masses from \co\ data. Envelope masses from the 1.3\,mm continuum are given in \cite{Stephens_2018}; note that these masses should be updated based on the revised distance estimate of Perseus of $\sim$300\,pc, where \citet{Stephens_2018} uses a distance of 235\,pc.
%as CO(2--1) and $^{13}$CO(2--1) tend to be confused with the large scale emission near the systemic velocities

We follow the same method as \cite{Goodman_1993} to calculate the velocity gradients for the \co\ envelopes. We used the clean mask velocity maps discussed above, and applied the \texttt{curve\_fit()} method from the Python package \texttt{scipy.optimize} to fit the velocity map with a simple 2-dimensional linear model:
\begin{equation} \label{eqn:1}
    v = v_0 + a \Delta \alpha + b \Delta \delta \quad,
\end{equation}
where $\Delta \alpha$ is the displacement in right ascension and $\Delta \delta$ the displacement in declination from the center of the envelope in the image \citep{Goodman_1993}. After fitting $a$ and $b$ for each source, we calculated the gradient $\mathcal{G}$ and gradient direction (measured east of north) $\theta_{\mathcal{G}}$:
\begin{equation} \label{eqn:2}
    \mathcal{G} \equiv (a^2 + b^2)^{1/2} \quad,
\end{equation}
\begin{equation} \label{eqn:3}
    \theta_{\mathcal{G}} = \arctan \frac{a}{b} \quad.
\end{equation}
The units for $a$ and $b$ are the same as that for the gradient, which for our study is km s$^{-1}$ pc$^{-1}$.

We used the multiplicity of the protostellar systems in order to classify the sources in our sample. We defined three classifications: single, close binary, and medium binary. A close binary is a system with two continuum sources detected by the VANDAM survey within 200 au of each other, a medium binary is a system with two sources greater than 200 au but closer than 3000 au from each other, and any protostars with no other protostars within 3000 au are considered single sources. Close binaries are approximate disk-scale binaries, while medium binaries share (or previously shared) a common envelope.

%%%
%%% TABLE 1
%%% table:table1
%%%
\begin{longrotatetable}
\begin{deluxetable*}{cccccccccccc} \label{table:table1}
\tablecaption{Protostellar Demographics}
\tablehead{
Source & \multicolumn{2}{c}{Observation Pointing Position} & \tbol\ & \lbol\ & Class & Mult. & $(F_{\text{int}})_{\text{C}^{18}\text{O}}$ & $(F_{\text{peak}})_{\text{C}^{18}\text{O}}$ & Axial & $\theta_{Env}$ & $\theta_{OF}$  \\
Name & R.A. (J2000) & Decl. (J2000) & [K] & [$L_\odot$] & & & [Jy km/s] & [Jy/bm km/s] & Ratio & [deg, E of N] & [deg, E of N]
}
\startdata
Per-emb-1   & 03:43:56.53               & +32:43:56.53              & $27 \pm 1$    & $1.80 \pm 0.10$   & 0     & S & $6.7 \pm 0.3$	& $3.33 \pm 0.13$  & $0.67 \pm 0.08$ & $128 \pm 9$    & $115 \pm 2$ \\
Per-emb-2   & 03:32:17.95               & +30:49:47.60              & $27 \pm 1$    & $0.90 \pm 0.07$   & 0     & C & $10.3 \pm 0.3$	& $2.25 \pm 0.09$  & $0.75 \pm 0.14$ & $119 \pm 22$   & $130 \pm 1$ \\
Per-emb-3   & 03:29:00.52               & +31:12:00.70              & $32 \pm 2$    & $0.50 \pm 0.06$   & 0     & S & $0.56 \pm 0.09$	& $0.45 \pm 0.06$  & $0.76 \pm 0.23$ & $90 \pm 51$    & $95 \pm 3$  \\
Per-emb-5   & 03:31:20.96               & +30:45:30.20              & $32 \pm 2$    & $1.30 \pm 0.10$   & 0     & C & $7.1 \pm 0.2$	& $2.03 \pm 0.07$  & $0.67 \pm 0.08$ & $101 \pm 8$    & $124 \pm 1$ \\
Per-emb-6   & 03:33:14.40               & +31:07:10.90              & $52 \pm 3$    & $0.30 \pm 0.00$   & 0     & S & $1.45 \pm 0.09$	& $0.25 \pm 0.03$  & $0.41 \pm 0.08$ & $159 \pm 5$    & $46 \pm 14$ \\
Per-emb-7   & 03:30:32.68               & +30:26:26.50              & $37 \pm 4$    & $0.15 \pm 0.06$   & 0     & S & $0.75 \pm 0.08$	& $0.28 \pm 0.03$  & $0.78 \pm 0.22$ & $107 \pm 59$   & $171 \pm 0$ \\
Per-emb-8   & 03:44:43.62               & +32:01:33.70              & $43 \pm 6$    & $2.60 \pm 0.50$   & 0     & S & $24.6 \pm 0.7$	& $3.17 \pm 0.14$  & $0.59 \pm 0.06$ & $90 \pm 5$     & $15 \pm 0$  \\
Per-emb-9   & 03:29:51.82               & +31:39:06.10              & $36 \pm 2$    & $0.60 \pm 0.06$   & 0     & S & $13.8 \pm 0.3$	& $2.23 \pm 0.06$  & $0.89 \pm 0.18$ & $57 \pm 60$    & $57 \pm 2$  \\
Per-emb-10  & 03:33:16.42               & +31:06:52.06              & $30 \pm 2$    & $0.60 \pm 0.05$   & 0     & S & $0.31 \pm 0.04$	& $0.14 \pm 0.02$  & $0.77 \pm 0.14$ & $159 \pm 23$   & $51 \pm 1$  \\
Per-emb-11  & 03:43:56.85               & +32:03:04.60              & $30 \pm 2$    & $1.50 \pm 0.10$   & 0     & M & $8.6 \pm 0.3$	& $0.95 \pm 0.07$  & $0.66 \pm 0.13$ & $171 \pm 12$   & $162 \pm 0$ \\
Per-emb-12  & 03:29:10.50               & +31:13:31.00              & $29 \pm 2$    & $7.00 \pm 0.70$   & 0     & M & $29.9 \pm 0.5$	& $5.64 \pm 0.10$  & $0.97 \pm 0.11$ & $12 \pm 86$    & $35 \pm 0$  \\
Per-emb-13  & 03:29:12.04               & +31:13:31.50              & $28 \pm 1$    & $4.00 \pm 0.30$   & 0     & S & $5.2 \pm 0.2$	& $2.00 \pm 0.06$  & $0.61 \pm 0.05$ & $0 \pm 4$      & $177 \pm 4$ \\
Per-emb-14  & 03:29:13.52               & +31:13:58.00              & $31 \pm 2$    & $0.7 \pm 0.08$    & 0     & S & $5.3 \pm 0.3$	& $1.27 \pm 0.09$  & $0.61 \pm 0.09$ & $104 \pm 10$   & --          \\
Per-emb-15  & 03:29:04.05               & +31:14:46.60              & $36 \pm 4$    & $0.40 \pm 0.10$   & 0     & S & $10.6 \pm 0.4$	& $1.81 \pm 0.11$  & $0.47 \pm 0.07$ & $97 \pm 5$     & $148 \pm 3$ \\
Per-emb-16  & 03:43:50.96               & +32:03:16.70              & $39 \pm 2$    & $0.40 \pm 0.04$   & 0     & S & $2.6 \pm 0.2$	& $0.57 \pm 0.05$  & $0.88 \pm 0.16$ & $160 \pm 38$   & $83 \pm 72$ \\
Per-emb-17  & 03:27:39.09               & +30:13:03.00              & $59 \pm 11$   & $4.20 \pm 0.10$   & 0     & C & $8.1 \pm 0.3$	& $0.56 \pm 0.06$  & $0.71 \pm 0.09$ & $3 \pm 10$     & $60 \pm 3$  \\
Per-emb-18  & 03:29:10.99               & +31:18:25.50              & $59 \pm 12$   & $2.80 \pm 1.70$   & 0     & C & $17.1 \pm 0.5$	& $1.92 \pm 0.11$  & $0.63 \pm 0.08$ & $115 \pm 8$    & $150 \pm 0$ \\
Per-emb-19  & 03:29:23.49               & +31:33:29.50              & $60 \pm 3$    & $0.36 \pm 0.05$   & 0/I   & S & $4.5 \pm 0.2$	& $0.96 \pm 0.06$  & $0.63 \pm 0.09$ & $75 \pm 9$     & $147 \pm 3$ \\
Per-emb-20  & 03:27:43.23               & +30:12:28.80              & $65 \pm 3$    & $1.40 \pm 0.20$   & 0/I   & S & $8.3 \pm 0.2$	& $0.85 \pm 0.05$  & $0.76 \pm 0.10$ & $170 \pm 15$   & $121 \pm 6$ \\
Per-emb-21  & \multicolumn{2}{c}{In same field as Per-emb-18}& $45 \pm 12$   & $6.90 \pm 1.90$   & 0     & S & $4.9 \pm 0.3$	& $1.92 \pm 0.11$  & $0.50 \pm 0.07$ & $55 \pm 5$     & $71 \pm 23$ \\
Per-emb-22  & 03:25:22.33               & +30:45:14.00              & $43 \pm 2$    & $3.60 \pm 0.50$   & 0     & M & $12.8 \pm 0.5$	& $3.39 \pm 0.13$  & $0.68 \pm 0.12$ & $29 \pm 15$    & $122 \pm 4$ \\
Per-emb-23  & 03:29:17.16               & +31:27:46.40              & $42 \pm 2$    & $0.80 \pm 0.10$   & 0     & S & $17.4 \pm 0.4$	& $1.28 \pm 0.06$  & $0.79 \pm 0.06$ & $90 \pm 10$    & $58 \pm 1$  \\
Per-emb-24  & 03:28:45.30               & +31:05:42.00              & $67 \pm 10$   & $0.43 \pm 0.01$   & 0/I   & S & $3.0 \pm 0.2$	& $0.41 \pm 0.04$  & $0.53 \pm 0.09$ & $14 \pm 7$     & $90 \pm 5$  \\
Per-emb-25  & 03:26:37.46               & +30:15:28.00              & $61 \pm 12$   & $1.20 \pm 0.02$   & 0/I   & S & $1.8 \pm 0.1$	& $0.41 \pm 0.05$  & $0.61 \pm 0.12$ & $10 \pm 11$    & $105 \pm 1$ \\
Per-emb-26  & 03:25:38.95               & +30:44:02.00              & $47 \pm 7$    & $8.40 \pm 1.50$   & 0     & M & $13.3 \pm 0.3$	& $1.77 \pm 0.06$  & $0.55 \pm 0.07$ & $147 \pm 5$    & $162 \pm 1$ \\
Per-emb-27  & 03:28:55.56               & +31:14:36.60              & $69 \pm 1$    & $19.00 \pm 0.40$  & 0/I   & C & $16.5 \pm 0.5$	& $2.73 \pm 0.10$  & $0.70 \pm 0.10$ & $7 \pm 11$     & $12 \pm 3$  \\
Per-emb-28  & \multicolumn{2}{c}{In same field as Per-emb-16}& $45 \pm 2$    & $0.70 \pm 0.08$   & 0     & S & $0.56 \pm 0.08$	& $0.57 \pm 0.05$  & $0.88 \pm 0.21$ & $159 \pm 54$   & $116 \pm 4$ \\
Per-emb-29  & 03:33:17.85               & +31:09:32.00              & $48 \pm 1$    & $3.70 \pm 0.40$   & 0     & S & $8.8 \pm 0.3$	& $3.20 \pm 0.10$  & $0.82 \pm 0.10$ & $157 \pm 27$   & $122 \pm 11$\\
Per-emb-30  & 03:33:27.28               & +31:07:10.20              & $78 \pm 6$    & $1.70 \pm 0.01$   & 0/I   & S & $7.5 \pm 0.3$	& $2.60 \pm 0.10$  & $0.36 \pm 0.07$ & $22 \pm 4$     & $122 \pm 0$ \\
Per-emb-32  & 03:44:02.40               & +32:02:04.90              & $57 \pm 10$   & $0.30 \pm 0.10$   & --    & M & $7.9 \pm 0.4$	& $0.94 \pm 0.10$  & $0.65 \pm 0.08$ & $101 \pm 8$    & --          \\
Per-emb-33  & 03:25:36.48               & +30:45:22.30              & $57 \pm 3$    & $8.30 \pm 0.80$   & 0     & C & $14.9 \pm 0.4$	& $2.15 \pm 0.10$  & $0.88 \pm 0.10$ & $31 \pm 26$    & $122 \pm 0$ \\
Per-emb-34  & 03:30:15.12               & +30:24:49.20              & $99 \pm 13$   & $1.60 \pm 0.10$   & I     & S & $4.3 \pm 0.2$	& $0.79 \pm 0.07$  & $0.56 \pm 0.08$ & $143 \pm 7$    & $58 \pm 0$  \\
Per-emb-35  & 03:28:37.09               & +31:13:30.70              & $103 \pm 26$  & $9.10 \pm 0.30$   & I     & M & $19.8 \pm 0.5$	& $2.69 \pm 0.09$  & $0.61 \pm 0.06$ & $57 \pm 6$     & $123 \pm 0$ \\
Per-emb-36  & 03:28:57.36               & +31:14:15.70              & $106 \pm 12$  & $5.30 \pm 1.00$   & I     & M & $28.1 \pm 0.9$	& $3.08 \pm 0.20$  & $0.67 \pm 0.06$ & $27 \pm 7$     & $23 \pm 2$  \\
Per-emb-38  & 03:32:29.18               & +31:02:40.90              & $115 \pm 21$  & $0.54 \pm 0.01$   & --    & S & $2.6 \pm 0.2$	& $0.77 \pm 0.08$  & $0.84 \pm 0.20$ & $144 \pm 57$   & --          \\
Per-emb-40  & 03:33:16.66               & +31:07:55.20              & $132 \pm 25$  & $1.50 \pm 1.00$   & I     & C & $10.1 \pm 0.6$	& $2.63 \pm 0.20$  & $0.50 \pm 0.08$ & $17 \pm 6$     & $106 \pm 5$ \\
Per-emb-42  & \multicolumn{2}{c}{In same field as Per-emb-26}& $163 \pm 51$  & $0.68 \pm 0.85$   & I     & M & $8.7 \pm 0.3$	& $0.94 \pm 0.06$  & $0.73 \pm 0.06$ & $142 \pm 8$    & $43 \pm 0$  \\
Per-emb-44  & 03:29:03.42               & +31:15:57.72              & $188 \pm 9$   & $32.50 \pm 7.10$  & 0/I   & M & $39.3 \pm 1.3$	& $7.37 \pm 0.32$  & $0.71 \pm 0.08$ & $39 \pm 9$     & $151 \pm 21$ \\
Per-emb-46  & 03:28:00.40               & +30:08:01.30              & $221 \pm 7$   & $0.30 \pm 0.07$   & I     & S & $0.24 \pm 0.07$	& $0.28 \pm 0.07$  & $0.83 \pm 0.33$ & $95 \pm 53$    & $133 \pm 2$ \\
Per-emb-47  & 03:28:34.50               & +31:00:51.10              & $230 \pm 17$  & $1.20 \pm 0.10$   & I     & S & $0.95 \pm 0.08$	& $0.26 \pm 0.03$  & $0.45 \pm 0.10$ & $107 \pm 7$    & --          \\
Per-emb-48  & 03:27:38.23               & +30:13:58.80              & $238 \pm 14$  & $0.87 \pm 0.04$   & I     & S & <0.11& --   & --    & --             & --          \\
Per-emb-49  & 03:29:12.94               & +31:18:14.40              & $239 \pm 68$  & $1.10 \pm 0.70$   & I     & S & <0.27& --   & --    & --             & $27 \pm 0$  \\
Per-emb-50  & 03:29:07.76               & +31:21:57.20              & $128 \pm 23$  & $23.2 \pm 3.0$    & I     & S & <0.42& --   & --    & --             & $99 \pm 6$  \\
Per-emb-52  & 03:28:39.72               & +31:17:31.90              & $278 \pm 119$ & $0.16 \pm 0.21$   & I     & S & <0.06& --   & --    & --             & $19 \pm 0$  \\
Per-emb-53  & 03:47:41.56               & +32:51:43.90              & $287 \pm 8$   & $4.70 \pm 0.90$   & I     & S & $13.4 \pm 0.4$	& $1.80 \pm 0.09$  & $0.67 \pm 0.07$ & $162 \pm 8$    & $59 \pm 0$  \\
Per-emb-54  & 03:29:01.57               & +31:20:20.70              & $131 \pm 63$  & $16.8 \pm 2.6$    & I     & S & $74.3 \pm 1.4$	& $4.79 \pm 0.22$  & $0.79 \pm 0.09$ & $5 \pm 15$     & --          \\
Per-emb-56  & 03:47:05.42               & +32:43:08.40              & $312 \pm 1$   & $0.54 \pm 0.09$   & I     & S & <0.21& --   & --    & --             & $147 \pm 2$ \\
Per-emb-57  & 03:29:20.07               & +31:23:14.60              & $313 \pm 200$ & $0.09 \pm 0.45$   & I     & S & <0.52& --   & --    & --             & $146 \pm 0$ \\
Per-emb-62  & 03:44:12.98               & +32:01:35.40              & $378 \pm 29$  & $1.80 \pm 0.40$   & I     & S & $0.37 \pm 0.06$	& $0.26 \pm 0.04$  & $0.56 \pm 0.15$ & $116 \pm 13$   & $21 \pm 3$  \\
Per-emb-63  & 03:28:43.28               & +31:17:33.00              & $436 \pm 9$   & $1.90 \pm 0.40$   & I     & S & <0.22& --   & --    & --             & --          \\
Per-emb-64  & 03:33:12.85               & +31:21:24.10              & $438 \pm 8$   & $3.20 \pm 0.60$   & I     & S & <0.23& --   & --    & --             & --          \\
Per-emb-65  & 03:28:56.31               & +31:22:27.80              & $440 \pm 191$ & $0.16 \pm 0.16$   & I     & S & <0.52& --   & --    & --             & --          \\
Per-emb-66  & 03:43:45.15               & +32:03:58.60              & $542 \pm 110$ & $0.69 \pm 0.22$   & I     & S & <0.15& --   & --    & --             & --          \\
SVS 13C     & 03:29:01.97               & +31:15:38.05              & $21 \pm 1$    & $1.50 \pm 0.20$   & 0     & S & $21.4 \pm 0.9$	& $4.49 \pm 0.26$  & $0.45 \pm 0.07$ & $93 \pm 5$     & $4 \pm 4$   \\
\enddata
\tablecomments{MASSES protostars and their bolometric temperatures and luminosities, class, total integrated ($F_{\text{int}}$) and peak integrated ($F_{\text{peak}}$) intensities, envelope elongation  $\theta_{Env}$, outflow axis $\theta_{OF}$, and axial ratios ($r_{min} / r_{maj}$) of the \co\ envelope. \tbol\ and \lbol\ values are from \cite{Tobin_2016}, and R.A., Decl., and class are from \cite{Stephens_2018}. Multiplicities are denoted as ``S" for single source, ``M" for medium binary, and ``C" for close binary.} Outflow axes from \citet{Stephens_2017} and Dunham et al. (in prep.). Errors for $F_{\text{int}}$ and $F_{\text{peak}}$ are derived from errors in moment 0 map; these do not account for absolute flux calibration errors, which we estimate to be $\sim$10-20\%.
\end{deluxetable*}
\end{longrotatetable}

\section{Analysis} \label{sec:analysis}

%In this section, we provide a demographic review of the protostellar envelopes from our analysis. The measured parameters for each protostar is given in Table~\ref{table:table1}. As will be shown, the histograms of most of the demographics are relatively uniform. Therefore, when we plot each distribution versus other characteristics of the protostellar system, we expect them to be statistically robust and not due to the underlying distributions of the plotted data.
To analyze how the \co\ envelopes evolve over time or as a result of other physical processes (e.g., outflows), we searched for relationships between physical parameters of the envelope and between the envelope parameters and parameters of other processes. As part of this analysis, we investigated how certain characteristics of the protostellar envelopes change as a function of \tbol. We further studied the velocity gradients and angular momenta of the \co\ envelopes and their dependency on the size of the envelope.

\begin{figure}
    \centering
    \includegraphics[width=\linewidth]{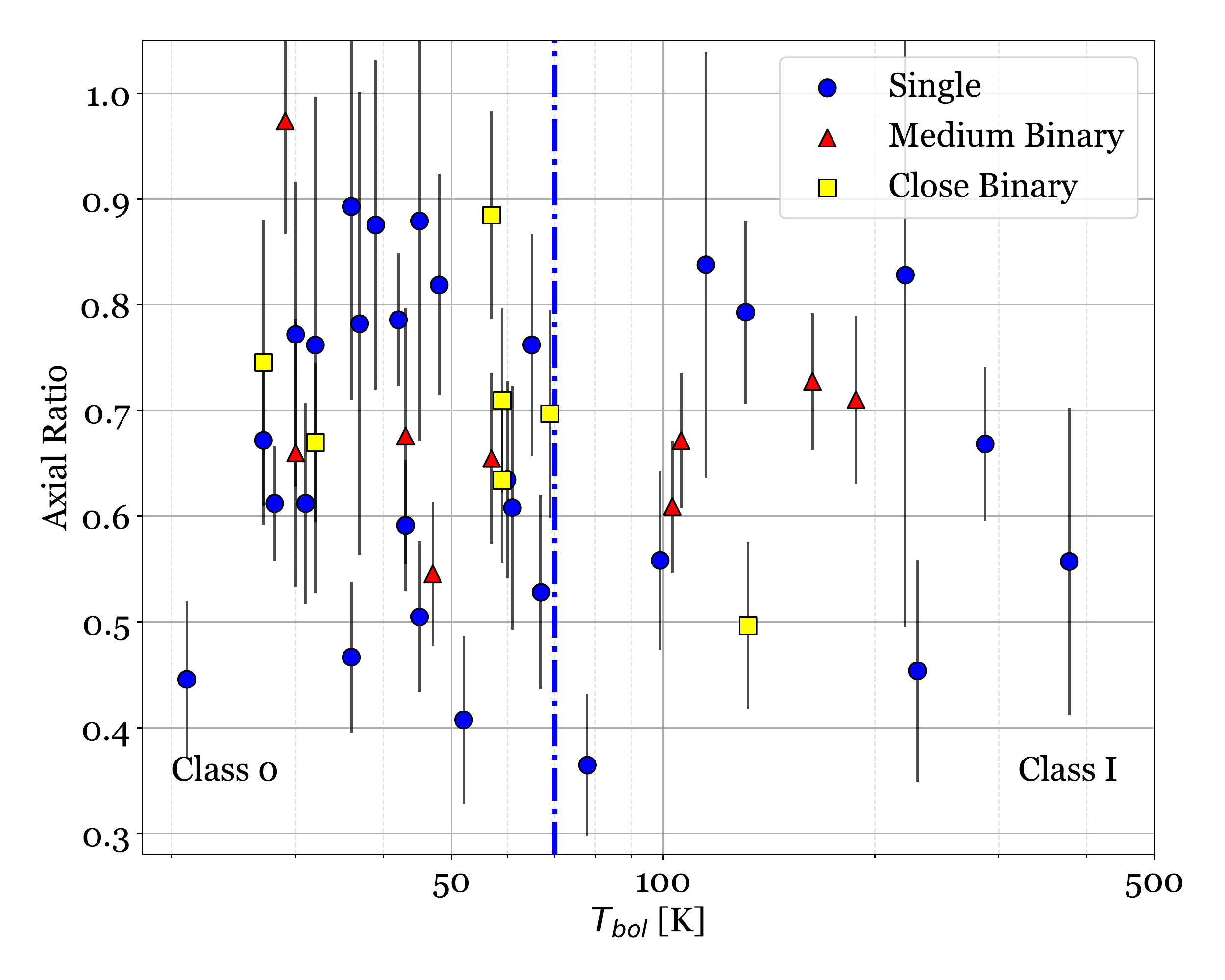}
    \caption{Axial ratios (between minor and major axes) of envelopes in relation to \tbol. The vertical blue line shows the approximate separation between Class 0 ($<$ 70 K) and Class I protostars; note Table~\ref{table:table1} shows more accurate classifications. All sources in Table \ref{table:table1} with a listed axial ratio are shown, which totals 26 C0, 11 CI, 7 C0/I.}
    \label{fig:axial_ratio_v_t}
\end{figure}

\subsection{Envelope Shapes, Intensities, and Velocity Gradients} \label{sec:shape}

In \S \ref{sec:intro}, we presented the empirical model for envelope evolution proposed by \cite{Arce_2006}. At early stages, the model predicts gas to be entrained along the outflows; nevertheless, it is possible the density of \co\ is too low to accurately trace this effect in some of these Class 0 sources. As the outflows widen and push the gas away from the outflow axis, we would expect Class I protostars to have \co\ envelopes that are elongated perpendicular to the outflow direction. We tested for this evolutionary trend by calculating the axial ratio and angular differences between a few key position angles, namely the \co\ elongation, the outflow axis direction, and the gradient direction. 

As mentioned in Section \ref{sec:methods}, we fit 2D Gaussian profiles to the integrated intensity (moment 0) maps and found the major and minor axes of the model for each source. Figure \ref{fig:axial_ratio_v_t} shows the distribution of axial ratios for the sources in relation to \tbol, where the error bars are from the elliptical Gaussian fits. (For Fig. \ref{fig:axial_ratio_v_t} and all subsequent figures, the number of sources of each class plotted is mentioned in the caption, where `C0' stands for Class 0, `CI' for Class I, and `C0/I' for Class 0/I.) No significant trends are found in axial ratio, although it should be noted that there are no high-\tbol\ protostars with near-circular envelopes (axial ratio > 0.85). The shapes of the envelopes can be visually seen in Appendix~\ref{appendix:A}, where the 2D Gaussian fits are overlaid on the moment 0 maps for each source. For several protostars, the \co\ moment 0 maps have envelope morphologies that are indicative of gas entrained along the outflow, especially those with an x-like morphology centered on the outflow axis (e.g., Per2, Per20, and Per35).

\begin{figure}
    \centering
    \includegraphics[width=\linewidth]{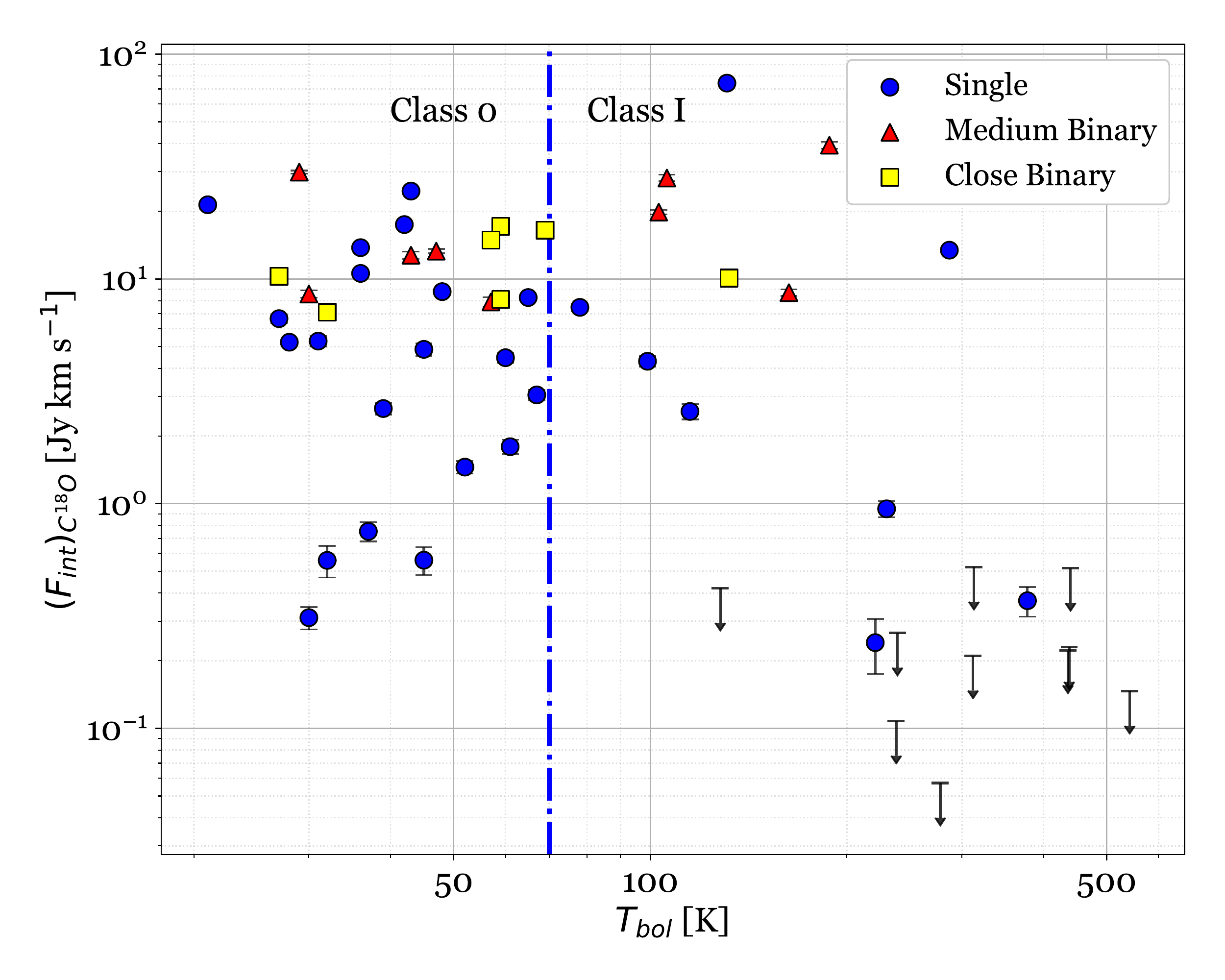}
    \caption{\co\ envelope total integrated flux versus \tbol\ for sources in the Perseus cloud studied in this paper, identified by color depending on the multiplicity of the system. We also found 10 sources that had continuum emission detected by the VANDAM survey but did not show any emission in \co; their 3-$\sigma$ upper bounds are shown (note that Per-emb-63 and 64 at a \tbol\ of $\sim$440~K overlap).} The vertical blue line shows the approximate separation between Class 0 ($<$ 70 K) and Class I protostars; note that Table~\ref{table:table1} shows more accurate classifications. All sources in Table \ref{table:table1} are plotted.
    \label{fig:flux_v_tbol}
\end{figure}

The total integrated flux over the solid angle of the source is vital information for understanding the relative amount of \co\ in the protostellar system. In \S \ref{sec:methods} we discussed the method used to calculate the total integrated flux of the envelopes. Figure \ref{fig:flux_v_tbol} shows the total integrated flux with respect to \tbol.

We detect a \co\ envelope for every Class 0 (as well as for every intermediate Class 0/I) source in our survey, but only detect envelopes in 13 of 23 Class I sources, for a $57 \pm 10$\% detection rate. This may indicate that some older protostars have already lost the majority of their natal envelopes. Note that these statistics exclude MASSES targets that were deemed to not be actual protostars (Per-emb-4, Per-emb-39, Per-emb-43, Per-emb-45, Per-emb-59, Per-emb-60, and Per-emb-61) and candidate first cores.

\begin{figure}
    \centering
    \includegraphics[width=\linewidth]{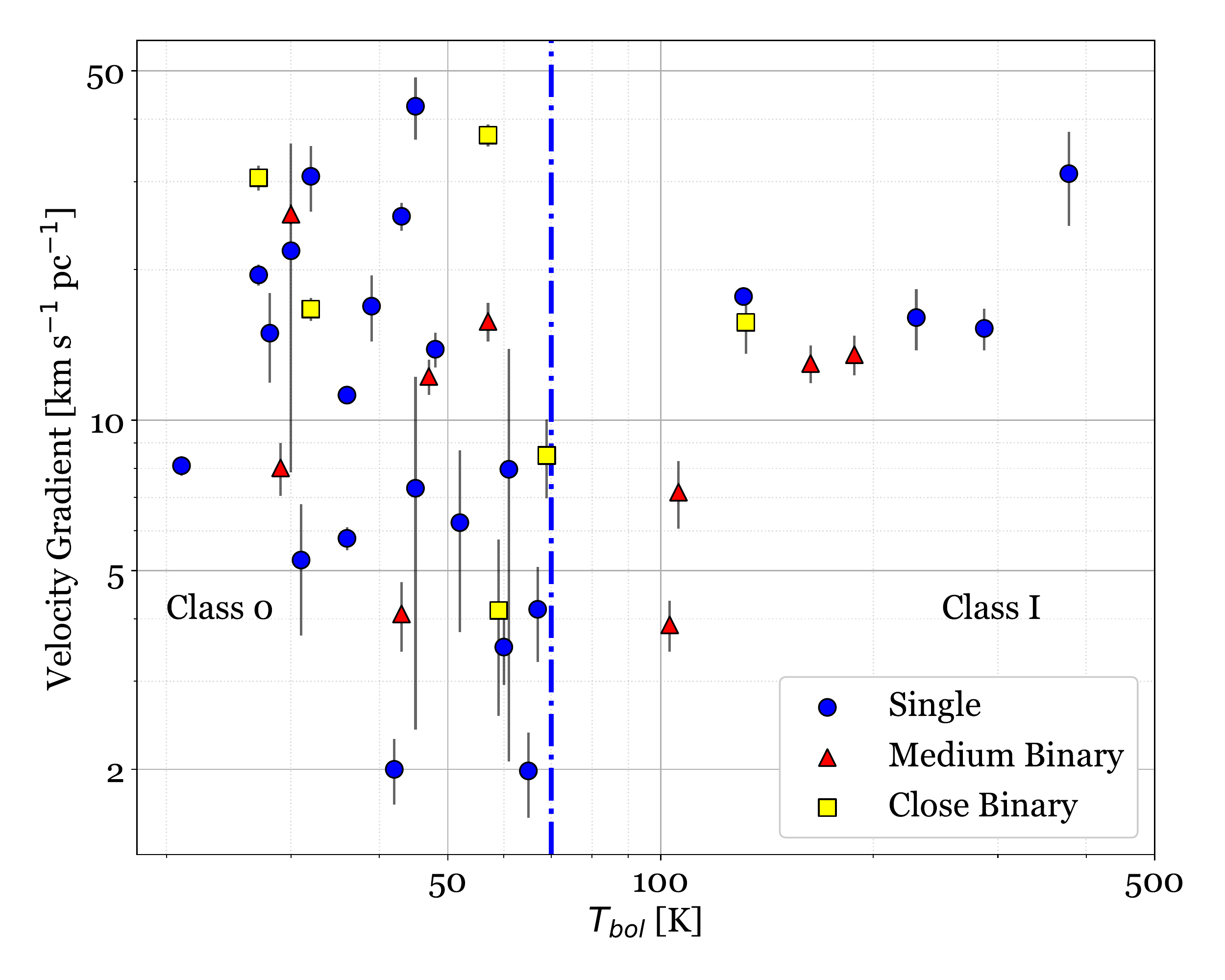}
    \caption{Velocity gradient of the \co\ envelopes versus \tbol.  The vertical blue line shows the approximate separation between Class 0 ($<$ 70 K) and Class I protostars; note that Table~\ref{table:table1} shows more accurate classifications. Plotted are all sources with a calculated velocity gradient in Table \ref{table:table2}, totaling 24 C0, 8 CI, and 6 C0/I sources.}
    \label{fig:tbol_v_grad}
\end{figure}

%One of the questions that we posed ourselves as we began to study the \co\ envelopes was ``how does the bipolar outflows affect the \co\ envelope?" 
In Section~\ref{sec:methods}, we discussed how we calculated the velocity gradients of the envelopes. Analyzing the velocity gradients of \co\ envelopes may give a good picture of how the dense gas is moving and how the outflows are affecting the flow of material in the system. (Note, the velocity maps and gradient fit are shown in Appendix~\ref{appendix:A}). Table \ref{table:table2} gives the velocity gradients, gradient directions, and axial ratios for sources with ``good" velocity gradient fits. Here ``good" fits are defined as those fits that both had a velocity map with a visually clear gradient across the envelope and had errors less than 50\% for both coefficients $a$ and $b$ in Eqn. \ref{eqn:1}. The velocity gradient values range from 2 km/s/pc up to 50 km/s/pc.

In Figure~\ref{fig:tbol_v_grad}, we plot the gradient with respect to \tbol, where the error bars come from the least squares fits of the velocity maps. No obvious relationship is found. Though not shown, we also looked at the relationship between the velocity gradient and the total integrated flux and the peak integrated flux. We did not find any significant trends, and there was no relationship between the gradient and flux and the multiplicity of the system. 

As shown, we did not find a significant relationship between integrated envelope \co\ flux and \tbol. Also, we did not find any relationship between the axial ratio of the \co\ envelope and \tbol\ (Figure~\ref{fig:axial_ratio_v_t}). This implies that regardless of mass, at young ages \co\ envelopes have a wide variety of shapes, ranging from nearly circular to having an axial ratio of over 2:1. The lack of a relationship appears to be independent of multiplicity.

\begin{figure*}
    \centering
    \includegraphics[width=0.80\textwidth]{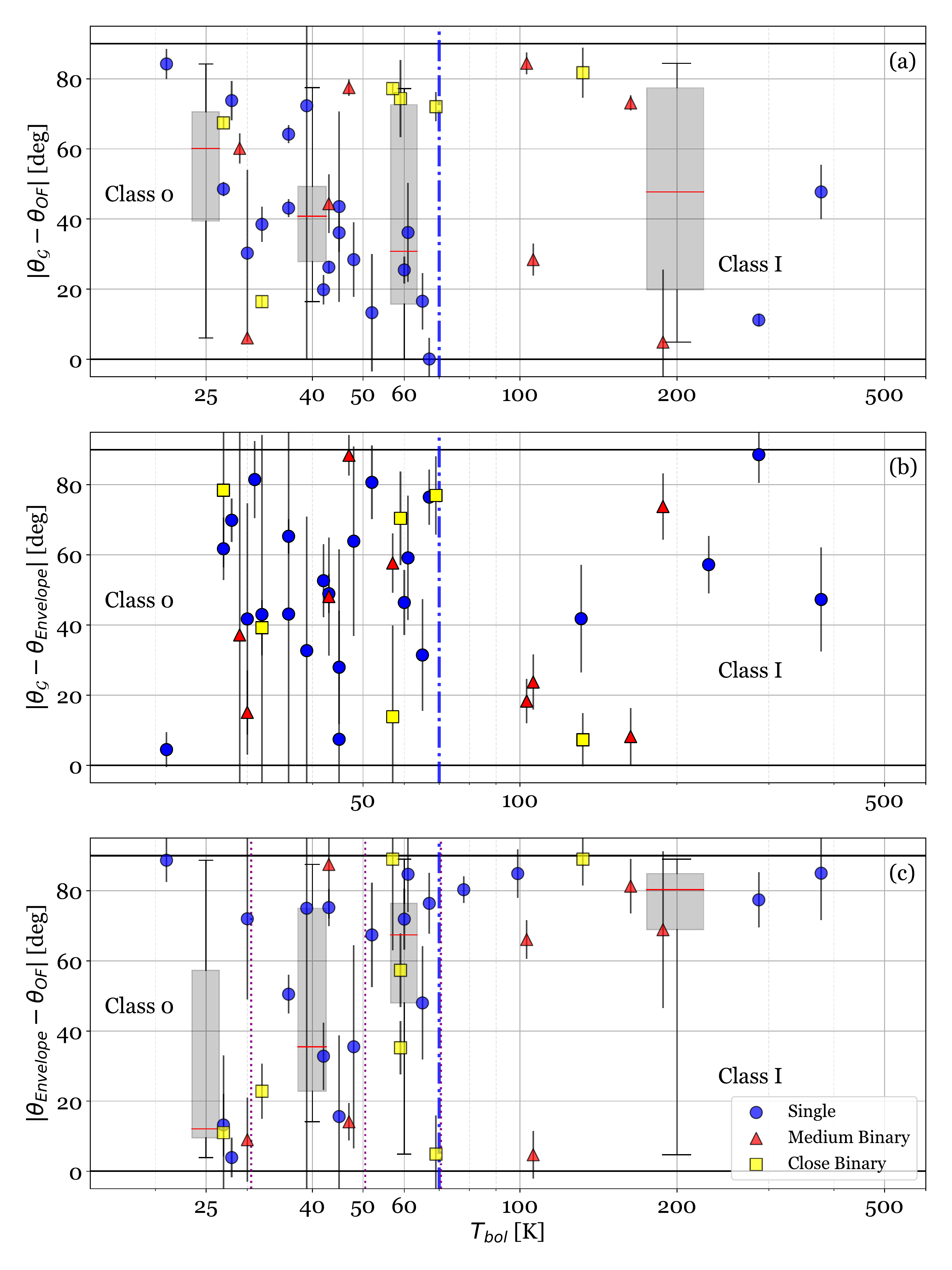}
    \caption{Angular differences between (a) the gradient direction and outflow axis direction, (b) the gradient direction and \co\ envelope major axis position angle, and (c) the envelope elongation position angle and outflow axis direction versus \tbol. The boxplots in (a) and (c) show the medians (red line), inter-quartile ranges (shaded region), and total ranges (black bars) of angle difference values for sources in the \tbol\ ranges (1) 10-30K, (2) 31-50K, (3) 51-70K, and (4) >70K. Purple dotted lines indicate the boundary of each \tbol\ range for the boxplots. Each point is color-coded based on the multiplicity of the protostars, and the legend in the bottom is for all panels. See \S \ref{sec:methods} for a definition of each multiplicity type.} The vertical blue line shows the approximate separation between Class 0 ($<$ 70 K) and Class I protostars; note that Table~\ref{table:table1} shows more accurate classifications. Each graph includes all sources that have both pertinent angles calculated (i.e. $\theta_{\text{Env}}$, $\theta_{\mathcal{G}}$, and $\theta_{\text{OF}}$) as found in Tables \ref{table:table1} and \ref{table:table2}, which totals (a) 22 C0, 6 CI, 6 C0/I, (b) 24 C0, 8 CI, 6 C0/I, (c) 19 C0, 8 CI, 6 C0/I.
    \label{fig:angle_comps}
\end{figure*}

%no longer analyzed
%We also studied the potential for the opening angle of the CO outflows to be another tracer of protostellar evolution. We would expect this to be true if the outflow model outlined in \cite{Arce_2006} was correct for every protostellar system.

\subsection{Envelope, Outflow, and Gradient Relative Orientations} \label{sec:temp_analysis}

Figure \ref{fig:angle_comps} plots against \tbol\ the angular difference between (a) the gradient direction ($\theta_{\mathcal{G}}$) and the outflow axis  ($\theta_{\text{OF}}$), (b) the gradient direction ($\theta_{\mathcal{G}}$) and the \co\ envelope elongation position angle ($\theta_{\text{Envelope}}$), and (c) the outflow axis ($\theta_{\text{OF}}$) and the \co\ envelope elongation ($\theta_{\text{Envelope}}$) (see Table \ref{table:table1} for values). The plots are color-coded based on the multiplicity of the system. Figure \ref{fig:angle_comps}(a) shows a possible trend of decreasing angle difference $|\theta_{\mathcal{G}}-\theta_{\text{OF}}|$ with increasing \tbol\ in Class~0 sources, but this disappears with the Class~I sample. As a test of this trend, we plotted box plots of the angle differences for sources in multiple \tbol\ ranges (10-30K, 31-50K, 51-70K, and $>$70K) on Figure \ref{fig:angle_comps}(a). While this hints at a weak trend for sources with \tbol$<$70K, there is not enough evidence to support any hypothesis of such a trend. For the $|\theta_{\mathcal{G}} - \theta_{\text{Envelope}}|$ vs \tbol\ panel (b), the early Class~I stages all have low absolute differences, which could indicate aligning of the gradient and envelope as protostars evolve into Class I stage. This trend disappears (or becomes harder to define) in the later Class I stage. Nevertheless, we do not have a large enough statistical sample to further investigate this possible trend.

\begin{figure}
    \centering
    \includegraphics[width=\linewidth]{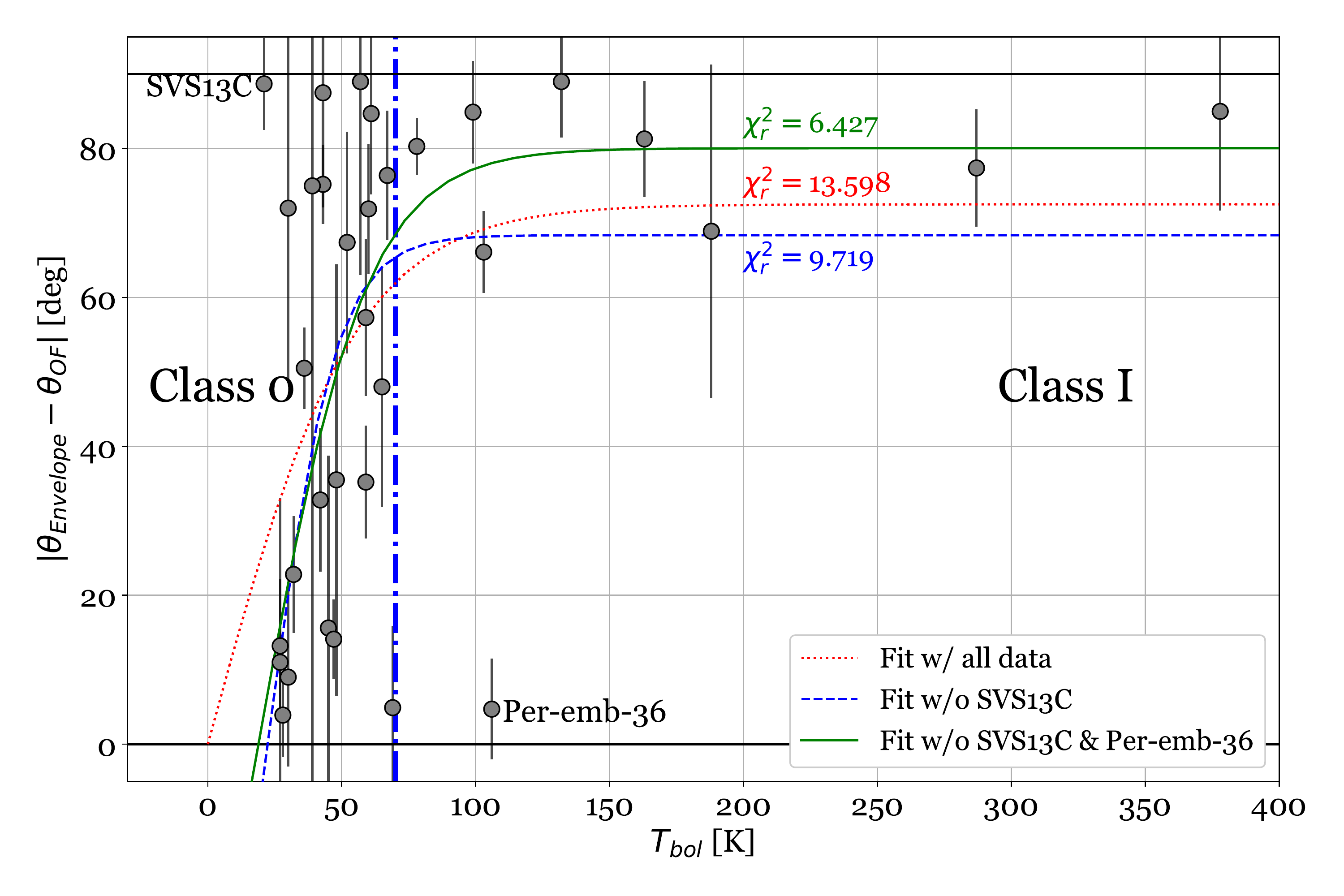}
    \caption{The difference in the elongation angle of the envelope and the position angle of the bipolar outflows versus \tbol. The data was fit with a tanh fit (see text) using all sources (red dotted), without Per-emb-36 (blue dashed), and without both Per-emb-36 and SVS13C (green solid). We find that with increased \tbol\ the envelopes become more perpendicular to the outflows (and hence more parallel with the accretion disks). With SVS13C and Per-emb-36 removed, we obtain values for the turnover \tbol\ and asymptotic angle difference more representative of the sample at large. The vertical blue line shows the approximate separation between Class 0 ($<$ 70 K) and Class I protostars; note that Table~\ref{table:table1} shows more accurate classifications. See Figure \ref{fig:angle_comps}(c) for number of sources plotted and their classes.}
    \label{fig:tbol_v_diff}
\end{figure}

A possible trend between $|\theta_{\text{Envelope}} - \theta_{\text{OF}}|$ and \tbol\ is evident in Figure \ref{fig:angle_comps}(c), with an increasing angular difference between the \co\ elongation and the outflow axis direction with increasing \tbol, during the Class 0 phase (\tbol\ $<$ 70 K), which flattens to an approximately constant value of $|\theta_{\text{Envelope}} - \theta_{\text{OF}}|$ for larger \tbol\ (Class I phase). To highlight this trend, we plotted the same boxplot analysis on \ref{fig:angle_comps}(c) as we did in panel (a) using the same \tbol\ ranges. With increasing \tbol, we find that $|\theta_{\text{Env}} - \theta_{\text{OF}}|$ increases towards 90 degrees, supporting the concept of flattening envelopes as the protostars evolve. The data shown in Figure \ref{fig:angle_comps}(c) are plotted again in Figure \ref{fig:tbol_v_diff} with a linear scale in \tbol\ for clarity of the subsequent model fitting.

To quantify the sharp change in envelope orientation, we need to fit a function that has a location where the curve turns over and asymptotes. One function that can be used is a tanh function:
\begin{equation}
    \Delta \theta_{\text{Env} - \text{OF}} = \theta_0 \tanh{\left(\frac{T - T_{\text{offset}}}{T_0}\right)}
\end{equation}
The turnover point of the fit can be estimated as the point with maximum curvature, $x_c$, i.e. where the function $f(x) = \tanh x$ maximizes the value of
\begin{equation}
    C(f) = f''(x)\left[ 1 + \left( f'(x) \right)^2 \right]^{-3/2}.
\end{equation}
For $f(x) = \tanh x$, the point of maximum curvature is at $x_c = 0.919$, which equates to a turnover temperature of $52 \pm 15$ K. This estimate requires a small correction, which must be computed numerically, so that $x_c$ is independent of changes in axis scaling. For the tanh function, this correction reduces $x_c$ by about 3\% \citep{Christopoulos2014}, which is negligible compared to our fit errors. We define $T_{\text{flex}}$ as the \tbol\ corresponding to $x_c$, i.e.,
\begin{equation}
    T_{\text{flex}} = 0.919 T_0 + T_{\text{offset}}\,.
\end{equation}
Using a tanh function is not unique, but we simply want a method that has a knee transition, as it is clear that the data make a sharp transition from low to high values at a \tbol\ between the typical value for class 0 and class I.

%and defined $T_{\text{flex}} = T_0 + T_{\text{offset}}$ as the \tbol\ at which the fit ``turns over" and asymptotes. 
We found a best fit of 
\begin{equation}
    \Delta \theta_{\text{Env} - \text{OF}} = (81 \pm 8)^{\circ} \tanh{\left[\frac{T_{\text{bol}}/\text{K} - (21 \pm 6)}{34 \pm 15}\right]}
\end{equation} 
when both SVS13C and Per-emb-36 were removed from the fit data, for a value of $T_{\text{flex}} = 53 \pm 20$ K. The fit without both of these points has the lowest reduced $\chi^2$ goodness-of-fit, although this is mostly due to the removal of the ``outlier" sources in the calculation of the test statistic. Here we chose to plot \tbol\ on a linear scale  to highlight the asymptotic behavior of the angle differences of the Class I sources. We will discuss the implication of this fit in \S \ref{sec:evolution_discussion}.
%These points are outliers to the fit, and are known to have complicated outflows \citep{Plunkett2013}

Because our data is skewed towards younger sources with lower \tbol, the above tanh fit may be similarly biased by the majority Class~0 population. To compensate for this, we used a bootstrapping method to iteratively select random samples of N Class~0 sources with replacement, where N is the total number of Class~I sources. (For this purpose, any sources classified as Class~0/I in Table \ref{table:table1} were classified by their \tbol\ value in relation to the separation value of \tbol\ = 70\,K.) We then fit the combined 2N Class~0 and I sources with the tanh model and calculated $T_{\text{flex}}$ for each fit. We found a $T_{\text{flex}}$ inter-quartile range of $[30, 72]$\,K and a median of 48\,K, which overlaps considerably with the error-bars for $T_{\text{flex}}$ calculated from the simple tanh fit described in the preceding paragraph.

We also color-code each target by multiplicity in Figures \ref{fig:angle_comps}. The relations for each panel in Fig.~\ref{fig:angle_comps} do not seem to be significantly different based on the multiplicity properties of each protostar.

\subsection{Multi-Scale Kinematics} \label{sec:mult_gradients}

In this subsection, we analyze MASSES velocity gradient data in the context of gradient analysis from other surveys of larger scale molecular line data. Specifically, we analyze velocity gradients and specific angular momenta with respect to the effective radii.

We calculated the effective radius \radeff, defined to be the radius of a circle with area equal to the projected area of the \co\ envelope. For this study, we took the boundary of an envelope to be at the 3-$\sigma$ contour of its moment 0 map. After tabulating this area $A$, we then calculated the radius by deconvolving the beam: 
\begin{equation} \label{eqn:4}
    R_{\text{eff}} = \sqrt{\frac{A - A_{\text{bm}}}{\pi}} ,
\end{equation}
where $A_{\text{bm}}$ is the area of the synthesized beam. Table \ref{table:table2} lists the effective radii for each envelope with a good gradient fit (see \S \ref{sec:shape} for definition of a ``good" fit). The radius values range from 0.002 to 0.011 pc, giving us about an order of magnitude range of radii.

\subsubsection{Velocity Gradient} \label{sec:gradients}
\begin{figure}
    \centering
    \includegraphics[width=\linewidth]{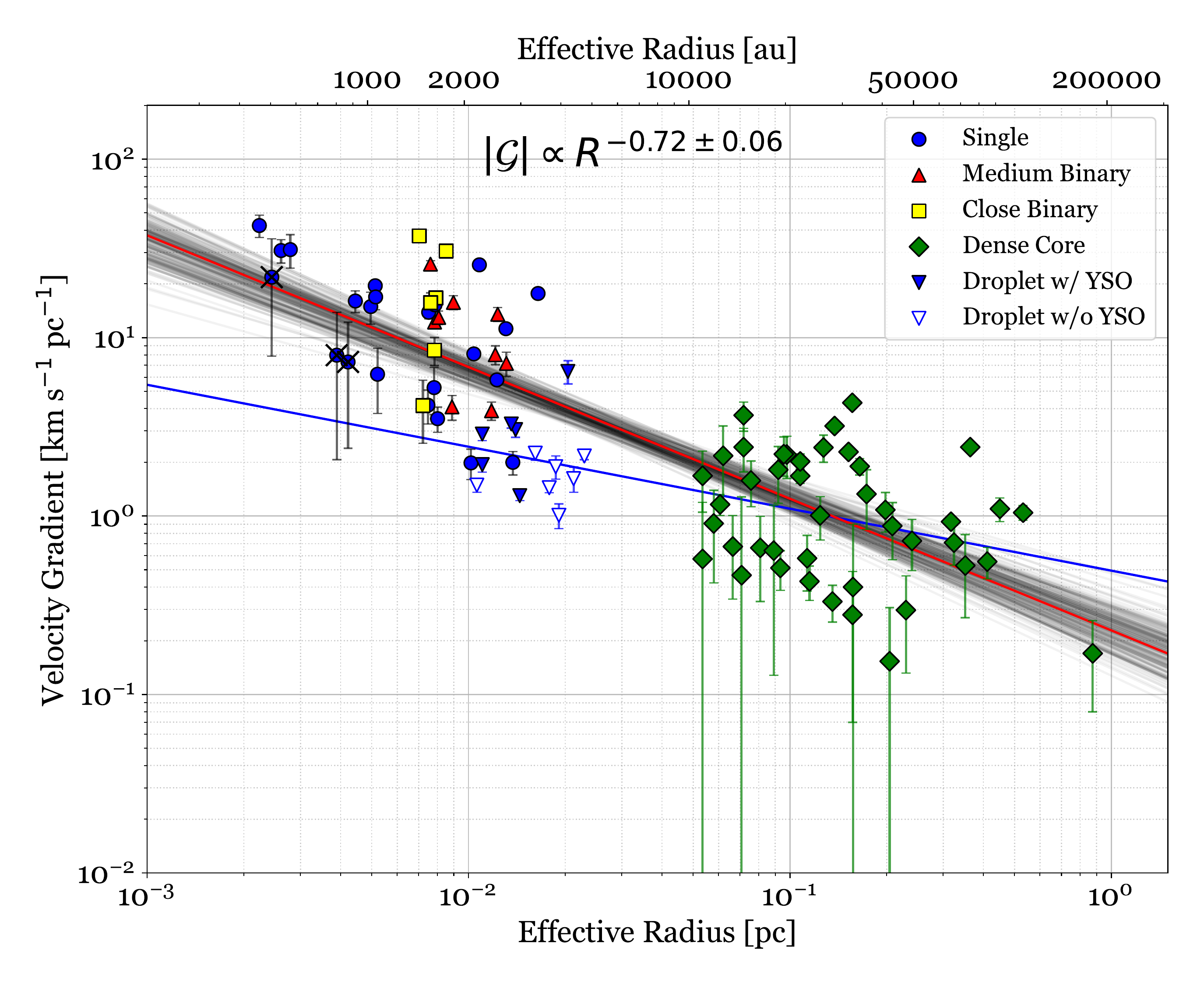}
    \caption{Velocity gradient versus effective radius for Perseus \co\ envelopes (solid blue circles, red triangles, yellow squares), droplets from \cite{Chen2019a} (blue solid and hollow triangles), and dense cores from \cite{Goodman_1993} (green diamonds). Black X's over markers indicate sources with velocity gradient uncertainty over 50\% (see Table \ref{table:table2}). The blue line is the best fit for the dense cores and droplets from \cite{Chen2019b} after dividing out Gaussian factor (see text). The red line is the best fit for all points, found using an MCMC sampling method. A random sample of 10\% of the fits calculated during the MCMC run are shown in gray as a rough indicator of fit uncertainty (only 10\% are shown for clarity). All sources from this study with calculated velocity gradient are included (see Table \ref{table:table2}), which amounts to 24 C0, 8 CI, and 6 C0/I sources.}
    \label{fig:grad_v_rad_mass}
\end{figure}

As mentioned above, we compare our measurements of the velocity gradient versus effective radius to those reported in the literature. Thus, we paired our sources with the dense cores studied in \cite{Goodman_1993} and the ``droplets" (sub-0.1 pc coherent gaseous structures) observed by the GAS collaboration in the L1688 region of Ophiuchus and the B18 region of Taurus in \cite{Chen2019a}. Combined with our data, we have a range of radii from 0.002 pc up to nearly 1 pc. While our sample includes only protostellar envelopes, the above two studies have a mix of protostellar and starless sources.
%We note that the methodology for calculating \radeff\ in these two studies differs from our study. \cite{Chen2019a} differs in that 
Figure~\ref{fig:grad_v_rad_mass} shows the relation between the velocity gradient and the effective radius for all of the sources listed above in the left plot along those with those from \citet{Goodman_1993} and \citet{Chen2019b}. We note that the distances for the \citet{Goodman_1993} sources have been updated with the more accurate measurements listed in \citet{Chen2019a}. We fit a simple power law model to the data using an unweighted MCMC sampling method from the Python package \texttt{pymc3}\footnote{See \href{https://docs.pymc.io/}{docs.pymc3.io}}, and found a relation of $|\mathcal{G}| \propto$~\radeff$^{-0.72 \pm 0.06}$. This power law is steeper than the fit found by \cite{Goodman_1993} of $|\mathcal{G}| \propto$~\radeff$^{-0.4 \pm 0.2}$ (dense cores only) and by \cite{Chen2019b} of $|\mathcal{G}| \propto$~\radeff$^{-0.45 \pm 0.13}$ (droplets and dense cores).

%Using the same analytical methods, we also fit a power law to the points in the right plot of Figure \ref{fig:grad_v_rad_mass}, which shows the velocity gradient versus the mass of the gaseous structures. To calculate the masses of the Perseus cloud \co\ envelopes, we used the local thermal equilibrium (LTE) mass using equation A9 in \cite{1997ApJ...476..781B}. We assumed an excitation temperature of $T_\text{ex} = 25$ K and a $[\text{H}_2/\text{C}^{18}\text{O}] = 5.88 \times 10^6$, as was done in \cite{Arce_2006}. The masses for the droplets were taken from \cite{Chen2019a}, and the masses for the dense cores were updated from \cite{Goodman_1993} using more accurate distances listed in Appendix A of \cite{Chen2019a}. We found a relation of $|\mathcal{G}| \propto M^{-0.38 \pm 0.03}$, significantly higher than the value found in (H. Chen et al., in prep) of $|\mathcal{G}| \propto M^{-0.20}$.

\begin{deluxetable}{ccccc} \label{table:table2}
%\tablecaption{Gradients, gradient directions, specific angular momenta, and effective radii for \co\ envelopes.}
\tablecaption{Envelope Gradients, gradient directions, specific angular momenta, and effective radii}
\tablehead{
Source & $|\mathcal{G}|$ & $\theta_{\mathcal{G}}$ & $J/M$ & \radeff \\
 & [km/s/pc] & [deg] & [10$^{-4}$ km/s pc] & [10$^{-3}$ pc]}
\startdata
Per-emb-1	& $19.5 \pm 0.9$	& $66 \pm 1$	& $1.3 \pm 0.1$	& 5.1 \\
Per-emb-2	& $30.5 \pm 1.7$	& $17 \pm 1$	& $5.6 \pm 0.3$	& 8.5 \\
Per-emb-3	& $30.7 \pm 4.6$	& $133 \pm 4$	& $0.5 \pm 0.1$	& 2.6 \\
Per-emb-5	& $16.7 \pm 0.9$	& $320 \pm 1$	& $2.6 \pm 0.1$	& 7.9 \\
Per-emb-6	& $6.2 \pm 2.5$	    & $59 \pm 9$	& $0.4 \pm 0.2$	& 5.2 \\
Per-emb-8	& $25.6 \pm 1.6$	& $221 \pm 2$	& $7.5 \pm 0.5$	& 10.8 \\
Per-emb-9	& $5.8 \pm 0.3$	    & $100 \pm 2$	& $2.2 \pm 0.1$	& 12.3 \\
Per-emb-10	& $21.8 \pm 13.9$	& $21 \pm 24$	& $0.3 \pm 0.2$	& 2.4 \\
Per-emb-11	& $25.8 \pm 1.2$	& $156 \pm 1$	& $3.7 \pm 0.2$	& 7.6 \\
Per-emb-12	& $8.0 \pm 1.0$	    & $335 \pm 4$	& $3.0 \pm 0.4$	& 12.1 \\
Per-emb-13	& $14.9 \pm 3.0$	& $250 \pm 4$	& $0.9 \pm 0.2$	& 5.0 \\
Per-emb-14	& $5.2 \pm 1.5$	    & $203 \pm 5$	& $0.8 \pm 0.2$	& 7.8 \\
Per-emb-15	& $11.2 \pm 0.3$	& $32 \pm 1$	& $4.8 \pm 0.1$	& 13.1 \\
Per-emb-16	& $16.9 \pm 2.6$	& $191 \pm 5$	& $1.1 \pm 0.2$	& 5.1 \\
Per-emb-18	& $4.2 \pm 1.6$	    & $225 \pm 10$	& $2.5 \pm 0.2$	& 7.2 \\
Per-emb-19	& $3.5 \pm 0.6$	    & $121 \pm 4$	& $0.6 \pm 0.1$	& 8.0 \\
Per-emb-20	& $2.0 \pm 0.4$	    & $138 \pm 5$	& $0.5 \pm 0.1$	& 10.2 \\
Per-emb-21	& $7.3 \pm 4.9$	    & $207 \pm 15$	& $0.3 \pm 0.2$	& 4.2 \\
Per-emb-22	& $4.1 \pm 0.7$	    & $77 \pm 8$	& $0.8 \pm 0.1$	& 8.9 \\
Per-emb-23	& $2.0 \pm 0.3$	    & $38 \pm 4$	& $0.9 \pm 0.1$	& 13.7 \\
Per-emb-24	& $4.2 \pm 0.9$	    & $90 \pm 3$	& $0.6 \pm 0.1$	& 7.5 \\
Per-emb-25	& $8.0 \pm 5.9$	    & $69 \pm 14$	& $0.3 \pm 0.2$	& 3.9 \\
Per-emb-26	& $12.2 \pm 1.0$	& $239 \pm 2$	& $3.9 \pm 0.3$	& 7.8 \\
Per-emb-27	& $8.5 \pm 1.5$ 	& $83 \pm 3$	& $1.3 \pm 0.2$	& 7.8 \\
Per-emb-28	& $42.5 \pm 6.0$	& $152 \pm 4$	& $0.5 \pm 0.1$	& 2.2 \\
Per-emb-29	& $13.9 \pm 1.1$	& $273 \pm 2$	& $2.0 \pm 0.2$	& 7.5 \\
Per-emb-32	& $15.7 \pm 1.4$	& $43 \pm 3$	& $3.2 \pm 0.3$	& 9.0 \\
Per-emb-33	& $37.2 \pm 1.9$	& $44 \pm 1$	& $4.6 \pm 0.2$	& 7.0 \\
Per-emb-35	& $3.9 \pm 0.5$	    & $219 \pm 3$	& $1.4 \pm 0.2$	& 11.8 \\
Per-emb-36	& $7.2 \pm 1.1$	    & $51 \pm 4$	& $3.1 \pm 0.5$	& 13.1 \\
Per-emb-40	& $15.7 \pm 2.1$	& $24 \pm 5$	& $2.3 \pm 0.3$	& 7.6 \\
Per-emb-42	& $13.0 \pm 1.1$	& $149 \pm 2$	& $4.2 \pm 0.4$	& 8.1 \\
Per-emb-44	& $13.5 \pm 1.2$	& $146 \pm 3$	& $5.1 \pm 0.5$	& 12.3 \\
Per-emb-47	& $16.0 \pm 2.2$	& $50 \pm 4$	& $0.8 \pm 0.1$	& 4.5 \\
Per-emb-53	& $15.3 \pm 1.5$	& $250 \pm 2$	& $2.4 \pm 0.2$	& 7.9 \\
Per-emb-54	& $17.7 \pm 0.6$	& $47 \pm 1$	& $12.0 \pm 0.4$& 16.5 \\
Per-emb-62	& $31.1 \pm 6.6$	& $69 \pm 7$	& $0.6 \pm 0.1$	& 2.8 \\
SVS13C	    & $8.1 \pm 0.4$	    & $268 \pm 2$	& $2.2 \pm 0.1$	& 10.4 \\
\enddata
\end{deluxetable}

We note that we modified the radius used in \cite{Chen2019a,Chen2019b}. The authors in these papers chose the droplet boundary using a similar method as we did for the envelope boundary (i.e., primarily using a cutoff based on the moment 0 signal-to-noise). They calculated the effective radii by using their boundary radii (referred to as ``brightness-weighted second moment" in \citealt{Chen2019a}) and multiplying them by a Gaussian factor of $2\sqrt{2\text{ln} 2}$. However, they only fit velocity gradient over their boundaries (as we did for the MASSES envelopes) and not over the extended area defined by the factor-multiplied radius. Therefore, to be consistent, we divide their radii by the Gaussian factor.  \cite{Goodman_1993} used a slightly different technique of estimating the effective radius and velocity gradients. Since their maps did not image the entirety of sources, they approximated the radius to be the geometric mean of the FWHM elliptical Gaussian fit to the major and minor axes, and fit velocity gradients over the entire maps. Given their different methodology, we do not modify their radii.

%We make two particular notes about this analysis. First, the method used to calculate the effective radius that we used is different than the method used in \cite{Chen2019a}. They defined the major axis $r_{\text{maj}}$ of a source as the FWHM along the direction of the highest dispersion in the second moment map using principal component analysis (PCA). Then, the minor axis $r_{\text{min}}$ is the FWHM of the second moment map in the direction perpendicular to $r_{\text{maj}}$. The effective radius is then defined as the geometric mean of the two axes, $R_\text{eff} = \sqrt{r_{\text{maj}}r_{\text{min}}}$. They note that this way of calculating \radeff\ only differs by at most 10\% from the method we used in this paper to calculate \radeff, which is within their errors. Thus, either method should produce similar results within the error bounds. While our analysis uses a 3$\sigma$ cut on the integrated intensity, we also repeated the analysis using 4$\sigma$ and 5$\sigma$ cuts, and recalculated the \radeff\ and $\mathcal{G}$. We found relationships consistent with our 3$\sigma$ cut, indicating that our results are not largely affected by the methodology in choosing \radeff.

We also note that the analyses in \cite{Goodman_1993} and \cite{Chen2019b} used the NH$_3$ line to calculate the gradients and radii, while we used the \co\ line. For the velocity gradient analysis, there should not be too large of a discrepancy between the results from using either spectral line, as they both should primarily trace the compact H$_2$ mass. 

\subsubsection{Specific Angular Momentum} \label{sec:spec_ang_mom}

A natural continuation of the analysis of velocity gradients is a study of the specific angular momentum of the \co\ envelopes. Previous studies \citep[e.g.,][]{Ohashi1997,Chen_2007,Belloche2013,Chen_2018,Pineda2019,Gaudel2020} showed that a relation exists between the specific angular momentum ($J/M$) and the effective radius (\radeff) of protostellar cores of $J/M \propto$ \radeff$^{q}$, where the power-law index $q$ is empirically measured with values of $\sim$1.5 to 1.8 for scales down to around 0.005 pc ($\sim$1000\,au). At scales smaller than $1000-2000$\,au, \citet{Ohashi1997} found that this relation flattens out (i.e., $q = 0$), which was recently confirmed by \citet{Gaudel2020} with a larger sample of sources. With our data, we were able to inspect the low end of this relation, reaching radii down to 0.0025\,pc ($\sim$500\,au). To derive $J/M$ for the envelopes, we used equation (5) of \cite{Chen_2007}:
\begin{equation} \label{eqn:5}
    J/M = \frac{2}{5 + 2p}\frac{\mathcal{G}}{\sin i} R^2 \approx \frac{1}{4}\mathcal{G} R^2
\end{equation}
where $p$ is the power law index of the radial density profile, $i$ the inclination angle of the line-of-sight to the rotation axis, and $\mathcal{G}$ the velocity gradient. Like \citet{Chen_2007}, we choose $p$~=~1.5 to be consistent with \citet{Goodman_1993}. This choice will allow us to better compare our results with those from \citet{Goodman_1993}. As with \cite{Chen_2007}, we assumed $\sin i = 1$. The $J/M$ values for the MASSES \co\ envelopes with calculated gradients are given in Table \ref{table:table2}.

\begin{figure}
    \centering
    \includegraphics[width=\linewidth]{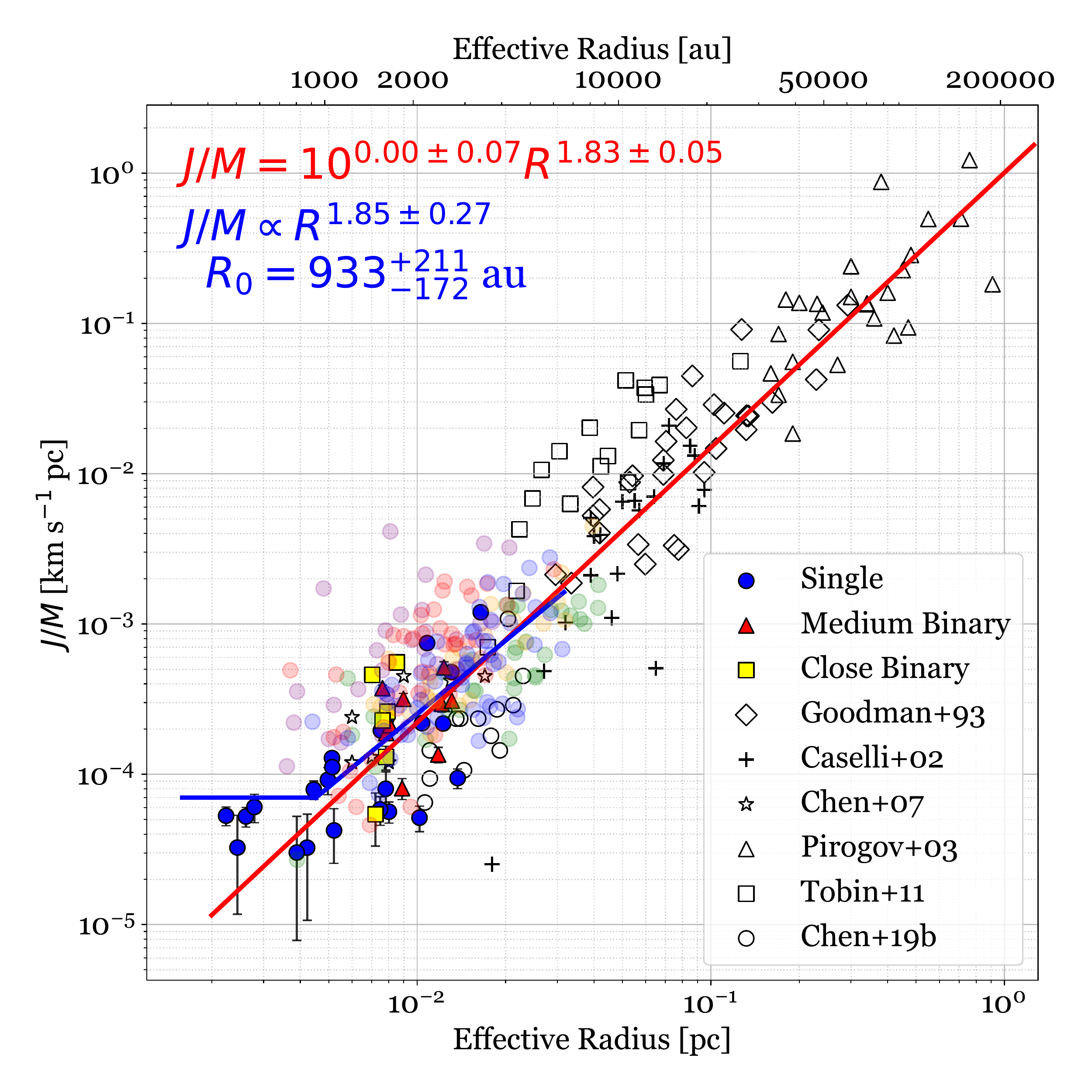}
    \caption{Specific angular momentum as a function of effective radius of the \co\ envelopes from this work (solid colors, color-coded by multiplicity), along with data from previous studies \citep{Goodman_1993,Caselli2002,Chen_2007,Pirogov2003,Tobin2011,Chen2019b} and simulation data from \citet{Chen_2018} (translucent circles, where each color shows a different data set). We fit a simple power law to all observed data (i.e. not simulated points), shown in red, and a composite power law with cutoff to envelope-scale data (this study, \citet{Chen_2007}, and \citet{Chen2019b}) shown in blue. These data alone do not indicate any definitive flattening in this relation; the composite fit simply shows that our data are not inconsistent with a flattening. All sources from this study with a calculated $J/M$ (see Table \ref{table:table2}) are plotted, which amounts to 24 C0, 8 CI, and 6 C0/I sources.}
    \label{fig:JM_v_rad}
\end{figure}

Figure~\ref{fig:JM_v_rad} shows the specific angular momentum ($J/M$) versus \radeff\ for the \co\ envelopes with calculated velocity gradients and sources from past work on angular momenta of protostellar systems: \citealt{1993ApJ...413L..47M, Caselli2002, Chen_2007, Pirogov2003, Tobin2011, Chen2019b}. On this plot, we also show the collection of results from \citet{Chen_2018}, which includes points from simulations as well as results from observational studies. The values of ($J/M$) versus \radeff\ for the MASSES \co\ envelopes match well with those from the \citet{Chen_2018} simulations. Using observational data only and the \texttt{scipy.optimize.curve\_fit} method, we find a power law relation of $J/M \propto R_{\text{eff}}^{1.83 \pm 0.05}$, which agrees with previous empirical results \citep[e.g.,][]{Pineda2019}.

We further looked at a possible cutoff of the power law at and below envelope scales ($\lesssim5000$ au), which has been seen in data by, for example, \citet{Ohashi1997}, \citet{Belloche2013}, and \citet{Gaudel2020}, where the specific angular momentum becomes constant. To test this in our data, we used a composite power law fit that becomes constant at small radii:
\begin{equation}
    J/M = C \left[ R_0^p + \frac{1}{2}\left(1 + \text{erf}\left(\frac{R - R_0}{\sigma}\right)\right)\left( R^p - R_0^p \right) \right]
\end{equation}
where $\sigma << 0.1$ is a constant, assumed small to keep the ``knee" of the fit sharp, and erf is the Gauss error function. We fit this power law to the same observational data as the simple power law and found the cutoff to be statistically similar, with an adjusted $R^2 = 0.920$ compared to an adjusted $R^2 = 0.917$ for the simple fit. To better estimate the cutoff radius, we fit just envelope-scale data, i.e. data from this study, \citet{Chen_2007}, and \citet{Chen2019b}. We found the cutoff radius to be $R_0 = 933_{-172}^{+211}$ au, with the fit shown in blue in Fig~\ref{fig:JM_v_rad}. We note however, that the data presented in this study certainly does not reflect a better fit than a pure power-law. The purpose of this composite fit shown here is to show that these data are not inconsistent with a possible flattening of this relation as shown in other studies \citep{Ohashi1997,Belloche2013,Gaudel2020}.

\section{Discussion} \label{sec:discussion}

\subsection{Evolution of \co\ Orientation} \label{sec:evolution_discussion}

The findings from the plot of $|\theta_\text{Envelope} - \theta_\text{OF}|$ versus \tbol\ in Figure \ref{fig:tbol_v_diff} support the evolutionary process proposed in \cite{Arce_2006}. As the outflow opening angle increases, the amount of \co\ along the outflow axis direction decreases, as most of the denser gas is pushed out and away by the outflows, creating cavities in the parent core. This forms a flattened envelope structure that is mostly perpendicular to the outflow axis, and hence we would expect the angular difference between the envelope elongation and the outflow axis direction to be close to 90\arcdeg\ at later ages. 

We should note that for low \tbol\ (i.e., younger sources), the distribution of angular differences between the envelope elongation and outflow axis direction is fairly random. This is somewhat expected since the shape of the envelope at early stages can be highly affected by the initial structure. 

Via our tanh fit to the data in Figure~\ref{fig:tbol_v_diff}, we estimated the parameter $T_{\text{flex}}$ (where the fit ``turns over" and asymptotes) to be 53~$\pm$~20\,K. This bolometric temperature is similar to the typical evolutionary separation of Class~0 and~I sources of 70\,K given in \cite{Chen1995}. Further, the Monte Carlo tanh fitting method resulted in a $T_{\text{flex}}$ quartile range of [30,72]\,K and median of 48\,K, which again contains the conventional 70\,K value for \tbol\ class separation. We should note, however, that the range of uncertainty for $T_{\text{flex}}$ contains much of the low-\tbol\ sources in our sample, which indicates that the tanh model fitting is not robust enough to conclude on its own a definitive estimate of a separation \tbol\ value for Class 0 and Class I sources. Nevertheless, coupled with the visible trend in increased $\Delta \theta_{\text{Env - OF}}$ with \tbol\ as seen in the boxplots in Fig. \ref{fig:angle_comps}(c), the tanh fit indicates an increasingly perpendicular envelope in relation to the outflows with increasing \tbol. Further study via this method and better measurements of \tbol\ for protostellar systems will potentially provide a more accurate estimate of a separation of Class 0 and Class I sources by \tbol.

%For example, while an initial spherical envelope may be elongated along the outflows, an initial spheroid with its major axis perpendicular to the outflows may not be pulled enough to elongate parallel to the outflows. 
For a source with an envelope that has an original spheroidal morphology with its major axis perpendicular to the outflow, it may be hard for the outflow to entrain enough gas along its axis to produce an envelope elongated along the outflow axis. As mentioned in Section~\ref{sec:shape}, while \co\ does appear to trace entrained gas for some outflows, due to its low abundance compared to $^{12}$CO along the outflow it may not be sensitive to the entrained gas for every outflow at the early stages. Thus, the orientation of the \co\ envelope axis may be weakly associated with the outflow in Class 0 envelopes.

In Figure~\ref{fig:axial_ratio_v_t}, we showed that although \co\ envelopes are typically elongated, there is no trend between their axial ratio and \tbol. Together with the results above, this says that despite elongation during Class 0 stage (e.g., due to collapse along magnetic field lines), the direction of the envelope's elongation can later evolve to be perpendicular to the molecular outflow.

It is important to note that our MASSES sample is inherently biased toward younger protostars, i.e. Class~0 and Class~I sources that still have an intact, detectable \co\ envelope. This places definitive limits on the timescales of protostellar evolution that we can study, namely any protostars that have shed their envelopes and moving into the Class~II phase of evolution (Figure~\ref{fig:arcecartoon}). Nevertheless, any study such as this one that is concerned with the dynamics, morphology, and evolution of protostellar envelopes will necessarily have to look at a sample that is younger than the general population of protostars, and this needs to be addressed. This bias does not affect the strength of the results in this paper, as we are not claiming any evolutionary trends beyond the confines of protostars with envelopes. As a test of our Class~0 and Class~I samples, we performed a 2-sample Kolmogorov-Smirnov test and found a KS statistic of 0.472, which rejects the null hypothesis of the two samples being drawn from the same distribution at a value of $\alpha = 0.15$.

\subsection{Gradient -- Size Relation} \label{sec:grad_size_discussion}

Intuitively, we expect the velocity gradient of a system to increase with a decrease in radius, as material falls in towards the center and picks up velocity. In Figure \ref{fig:grad_v_rad_mass}, we plot the velocity gradient versus the effective radius for MASSES envelopes, and data presented in \cite{Goodman_1993} and \cite{Chen2019a,Chen2019b}. The inverse relation between gradient and size is shown to many orders of magnitude of radii for different gaseous structures. %We note that we modified the radius used in \cite{Chen2019a,Chen2019b}. The authors in these papers chose the droplet boundary in a similar method as we did for an envelope boundary (i.e., primarily using a cutoff based on the moment 0 signal-to-noise), but they multiplied their radii by a Gaussian factor of $2\sqrt{2\text{ln} 2}$. However, they only fit velocity gradient over their boundaries (as we did for the MASSES envelopes) and not over the extended area defined by the factor-multiplied radius. Therefore, to be consistent, we divide their radii by the Gaussian factor.  \cite{Goodman_1993} used a slightly different technique of estimating the effective radius and velocity gradients. Since their maps did not image the entirety of sources, they approximated the radius to be the geometric mean of the FWHM elliptical Gaussian fit to the major and minor axes, and fit velocity gradients over the entire maps. Given their different methodology, we do not modify their radii.

Compared to the relationship found in \citet{Goodman_1993} and \citet{Chen2019b}, we found a significantly higher slope, $|\mathcal{G}| \propto R_\text{eff}^{-0.72 \pm 0.06}$. \cite{Goodman_1993} found  the gradient goes as $R_\text{eff}^{-0.4}$ while \citet{Chen2019b} (combining data from \citealt{Goodman_1993}) found $R_\text{eff}^{-0.45}$ (if one does not correct the radius as we did above). The difference in these previous studies and our model fit is the availability of a larger sample size, resulting in a wider range of values for \radeff\ for fitting. 

The relationship also does not appear to be greatly affected by multiplicity. However, we do note that all the close binaries tend to lie above our best fit, indicating that close multiples may have higher velocity gradients than the simple trend indicates.

One question that might come naturally from this analysis is what this power law fit physically means. Should conservation of angular momentum hold, we would expect that as an envelope increases in radius, its velocity would decrease as $R^{-1}$. However, this assumes solid-body rotation, which is a very crude approximation for the turbulent, accreting envelope environment. Our result is not consistent with solid-body rotation, and we do not observe conservation of angular momentum over these size scales. Since our result of $R^{-0.72}$ is shallower than the $R^{-1}$ relation, this suggests a dissipation of angular momentum as material falls in towards the protostellar disk. There are many physical processes that dissipate angular momentum from the system, such as bipolar outflows and tension in the magnetic field lines. Further, gas is constantly being accreted onto the disk and being pulled into the envelope from the larger cores, and even the cores are accreting gas from the molecular cloud while the outflows are pushing gas out of the system. Thus, it is very difficult to understand a gaseous structure as its own entity without considering the processes at larger and smaller scales that affect it. 

It is also important to note that although the data appear to clearly follow a power law fit, we do not claim that our fit (or any power law, for that matter) accurately describes the physics happening in the protostellar systems. In fact, we would expect different efficiencies in angular momentum conservation for different length scales as different processes become more dominant. For example, while bipolar outflows could potentially disperse gas traced by \co\ and affect the calculated velocity gradients at the envelope scale, they probably do not grossly affect the molecular cloud ($>$10 pc scale).

Rather than measuring a single velocity gradient value for each protostar at an effective radius as we did in this paper, \cite{Gaudel2020} calculated velocity gradients at multiple radii from 50 to 5000\,au and fit this relationship for each of their 12 protostellar envelopes. In their results, the relationships between the velocity gradient and radius varies from protostar to protostar, but the median velocity gradient is proportional to radius to the --0.85 power. This value is similar to the $R_\text{eff}^{-0.72 \pm 0.06}$ relation we find. However, we note that the power-law index measured by \cite{Gaudel2020} varies markedly from protostar to protostar.

\begin{comment}
\begin{figure}
    \centering
    \includegraphics[width=\linewidth]{opening_v_ell_ang_out_ang.png}
    \caption{Angular difference between the \co\ envelope elongation angle and the outflow axis direction versus the outflow opening angle. While the plot has a similar shape to the analogous plot for \tbol\ (the lower plot of Figure \ref{fig:angle_comps}), we don't have enough statistics for high opening angles to deduce any relationship.}
    \label{fig:opening_v_ell_out}
\end{figure}
\end{comment}

\subsection{Specific Angular Momentum -- Size Relation}

In contrast to the velocity gradient, we expect the specific angular momentum ($J/M$) to increase with increasing radius, recalling Equation~\ref{eqn:5}: $J/M \propto |\mathcal{G}| R^2$. Noting the power law definition of $J/M$ in terms of \radeff, then, it is no surprise and in fact expected that we found a power law relationship in Figure~\ref{fig:JM_v_rad}. As mentioned before, several previous studies find power-law indices for $J/M \propto R^{q}$ to be between 1.5 and 1.8.
%Previous studies \citep[e.g.,][]{Chen2019b,Burkert2000} have found relations of $J/M \propto R^{1.5}$, which is consistent with the interpretation that the observed specific angular momentum is due to a turbulence scaling law with the canonical velocity dispersion $\sigma \propto R^{0.5}$.

Our analysis, coupled with past results, allows us to study this relationship at a wider range of core sizes, from 1 pc down to $\sim$0.0025\,pc (500\,au). We find a power law fit of $J/M \propto R^{1.83 \pm 0.05}$ (see Figure~\ref{fig:JM_v_rad}), which lies between the relations for solid-body rotation ($J/M \propto R^2$) and for turbulence ($J/M \propto R^{1.5}$) \citep{Burkert2000}. This is to be expected, as the envelope gas accrues the angular momentum of the infalling gas from the core, while processes such as  bipolar outflows and infalling gas onto the disk provide turbulence in the system.

Past papers, including \cite{Belloche2013}, have posited that at scales smaller than $\sim$5000 au, the specific angular momentum will be constant. This turnover is necessary to couple the turbulent dynamics of the core and envelope with the flat inner disk. Indeed, \citet{Gaudel2020} seems to find this turnover at around $\sim$1000\,au. Conversely, \cite{Pineda2019} did not find a break in the power law fits for any of their 3 sources' radial $J/M$ profiles down to 1000\,au. Our data, which contains 7 sources with $R_{\text{eff}} < 1000$\,au and 25 sources with $R_{\text{eff}} < 2000$\,au, cannot rule out such a flattening occurring in the region of $\sim$1000 au and consequently cannot confirm nor refute such a break in the data. As mentioned in Section~\ref{sec:spec_ang_mom}, this leveling out of the specific angular momentum as seen in Figure \ref{fig:JM_v_rad} is statistically similar in our data to a simple power law relation.

\section{Summary} \label{sec:conclusion}

Our analysis of 54 \co\ envelopes of protostars in the Perseus molecular cloud from the MASSES survey has characterized the shapes, sizes, and orientations of these intermediate-sized gaseous structures. We considered the effect of the bipolar outflows on the envelopes, and we measured the velocity gradients and specific angular momenta and compared them to larger cores in other molecular clouds from previous studies.
 
Comparing the elongation angle of the \co\ envelope with the direction of the bipolar outflows, we found that as the protostar evolves and the outflows widen over time (using \tbol\ as a tracer of protostellar age), the envelopes become flattened in the direction perpendicular to the outflows, supporting the picture of protostellar envelope evolution painted by \cite{Arce_2006}. This flattening occurs around a \tbol\ of $53 \pm 20$\,K, which includes the conventional separation of Class~0 and I protostellar stages of 70\,K \citep{Chen_2007}. Via bootstrapping with replacement, we found this relation holds even if we had an equal number of Class~0 than Class~I in our sample. However, we note that the lower error-bar encompasses almost all our data, so separation of the protstellar classifications via this method is tentative with these data.

%, but we also considered the opening angle of the outflows as another possible tracer of age. The angle difference between the elongation angle and the outflow axis direction compared to the opening angle of the outflows yielded similar trends at small opening angle ($\theta_\text{Open} < 80\deg$), but we did not have enough statistics at higher opening angles to take the opening angle as a tracer of age. 

We compared the velocity gradients of the envelopes to their effective radii, and using data from \cite{Goodman_1993} and \citet{Chen2019b} to increase the range of size scales up to 1 pc we found a power law behavior of $|\mathcal{G}| \propto R_\text{eff}^{-0.72 \pm 0.06}$. This is sufficiently different from a value of -1 that angular momentum is not conserved over the size scales investigated. We calculated the specific angular momenta of the protostellar envelopes using the velocity gradients, and using data from previous studies of larger gaseous protostellar structures, we found a power law relation of $J/M \propto R_\text{eff}^{1.83 \pm 0.05}$. Our data is not inconsistent with a turnover in the specific angular momentum at $\sim$1000\,au, with the value remaining constant at smaller radii, but this will require more data and further study to make a confident assertion.

We also investigated whether the underlying multiplicity affected any of these relationships. We did not see any changes in any of the relationships investigated, but we do note our statistics are somewhat limited when separating sources based on their multiplicity. 

\section*{Acknowledgements} \label{sec:acknowledgements}

D.J.H. acknowledges support from a Yale-Smithsonian Partnership fellowship. 
I.W.S. acknowledges support from NASA grant NNX14AG96G. 
H.G.A. acknowledges support from the National Science Foundation award AST-1714710.
I.W.S. and T.L.B. acknowledges support from the SMA and the Center for Astrophysics $|$ Harvard \& Smithsonian’s Radio and Geoastronomy Division.
The Submillimeter Array is a joint project between the Smithsonian Astrophysical Observatory and the Academia Sinica Institute of Astronomy and Astrophysics and is funded by the Smithsonian Institution and the Academia Sinica.
We would like to send thanks to Hope How-Huan Chen for providing data from the Droplets papers \citep{Chen2019a, Chen2019b} for the multi-scale kinematics analysis, and to Che-Yu Chen for providing simulation and observational data from \citet{Chen_2018}.

\software{CASA (McMullin et al. 2007), astropy (The Astropy Collaboration 2013, 2018), scipy (Jones et al. 2001), scikit-learn (Pedregosa et al. 2011), pymc3 (Salvatier, Wiecki \& Fonnesbeck 2016)}


\appendix \section{Moment 0, Velocity, \& Line Width Maps} \label{appendix:A}
\begin{figure*}
    \centering
    \includegraphics[height=0.93\textheight, keepaspectratio]{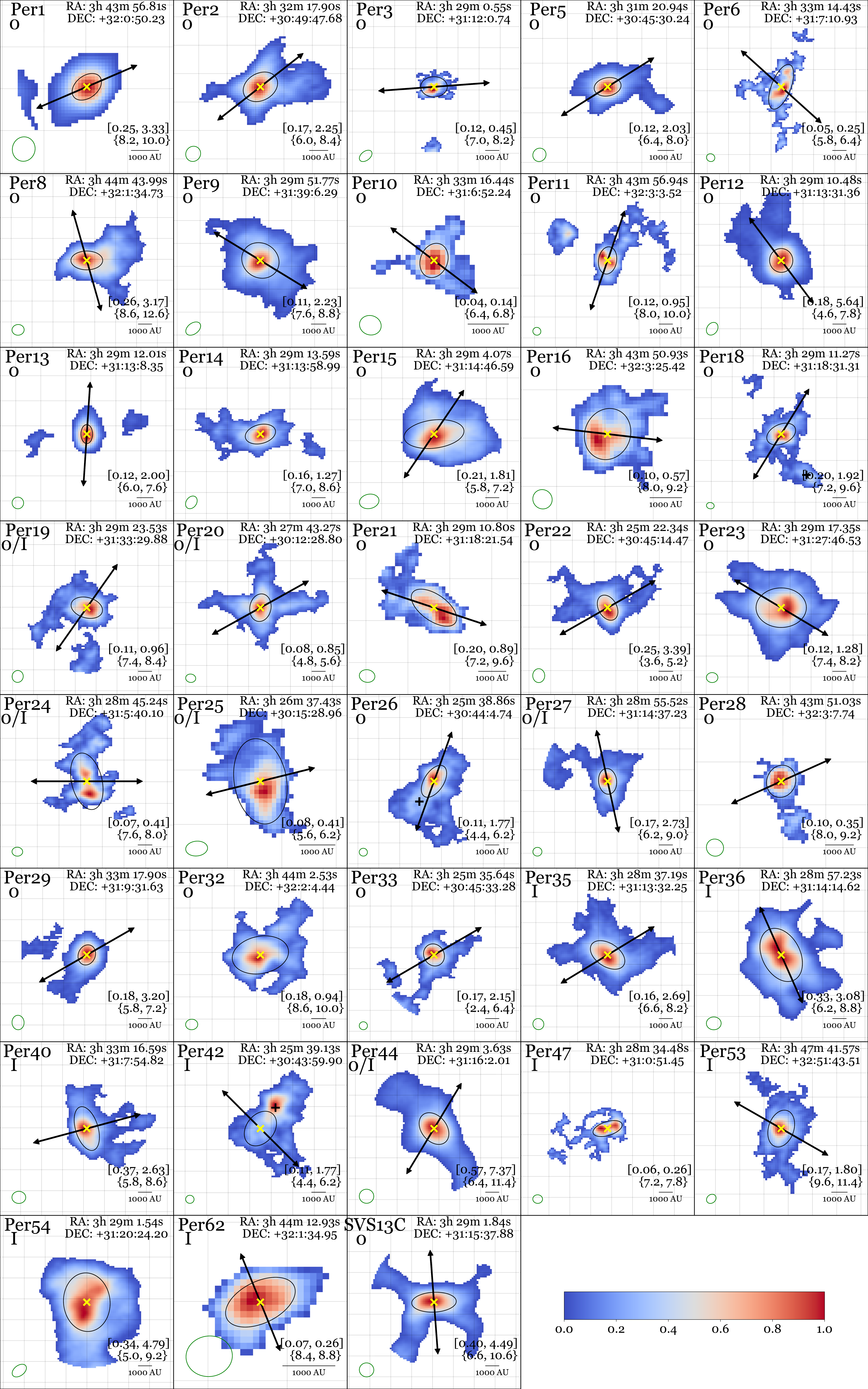}
    \caption{Moment 0 maps with a 2-$\sigma$ cutoff of MASSES survey \co\ envelopes studied in this paper. The large black ellipses are the FWHM of the 2D Gaussian fits on the moment 0 maps. The black arrows are the outflow direction and are centered on the center of the Gaussian fit ellipse, shown as a yellow X corresponding to the coordinates given at the top of the panel. Black +'s denote other protostars in the image (note that Per-emb-42 shares an apparent envelope with Per-emb-26, which has a higher intensity envelope).} The velocity range for the integration is given in curly brackets, in km s$^{-1}$, and the minimum and maximum pixel values are given in square brackets, in Jy/beam km/s. A normalized colorbar is given in the bottom right of the figure. The synthesized beam is in the bottom left in green. The class of each object is given below its name in the top left. All grids have 5" by 5" spacing, with a scalebar indicating 1000 au in the lower right corner of each plot.
    \label{fig:gallery_mom0}
\end{figure*}

\begin{figure*}
    \centering
    \includegraphics[height=0.95\textheight, keepaspectratio]{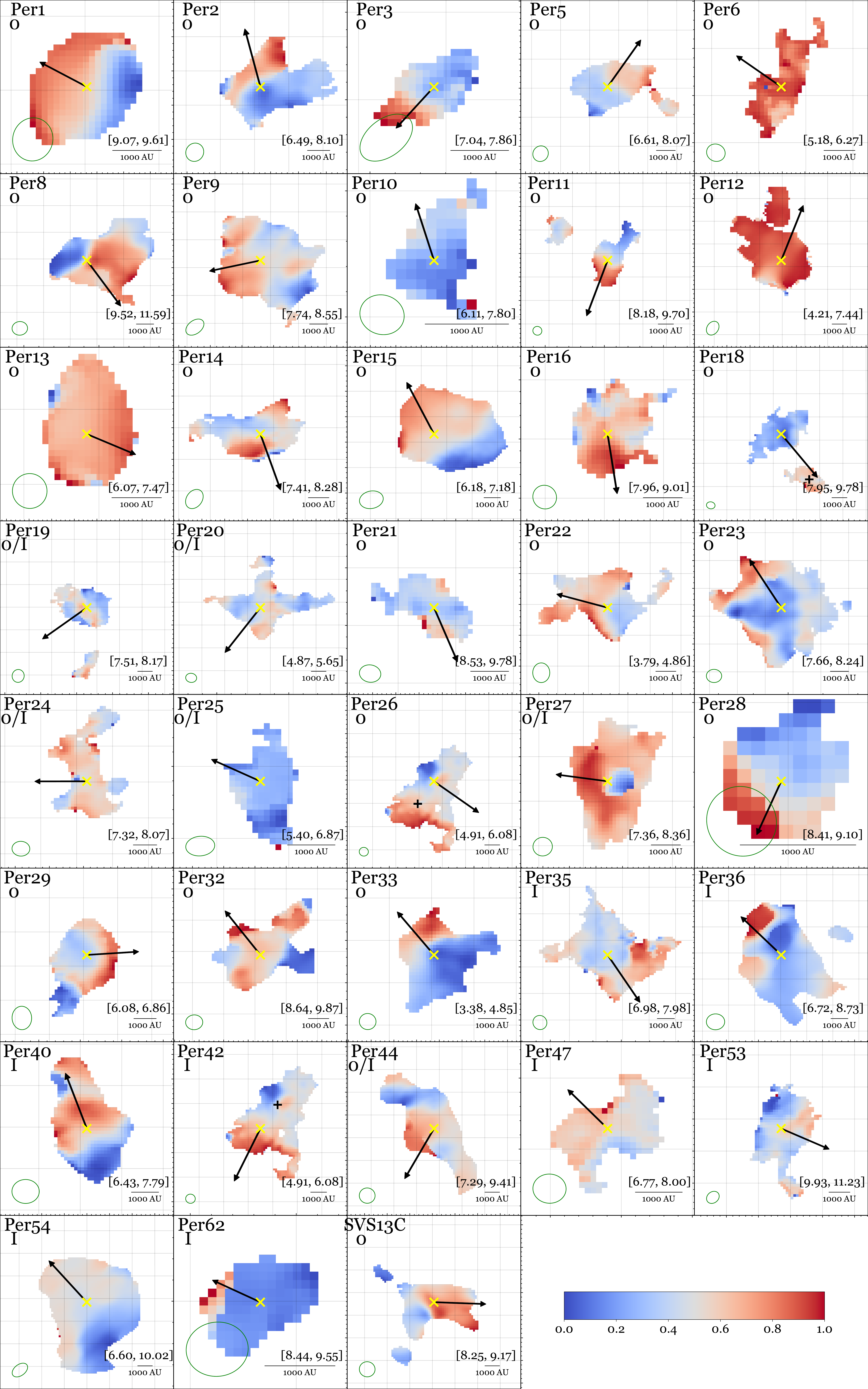}
    \caption{Velocity maps of MASSES survey sources studied in this paper. The black arrows are the angles of the gradient, pointed towards higher velocity. The minimum (blue) and maximum (red) velocities are given in brackets in km s$^{-1}$, with a linear color scale. The center of the 2D Gaussian fit (see Figure \ref{fig:gallery_mom0}) is shown as a yellow X, with black +'s denoting other sources in the map. The synthesized beam is in the lower left in green. All grids have 5" by 5" spacing, with a scalebar indicating 1000 au in the lower right corner of each plot.}
    \label{fig:gallery_mom1}
\end{figure*}

\begin{figure*}
    \centering
    \includegraphics[height=0.95\textheight, keepaspectratio]{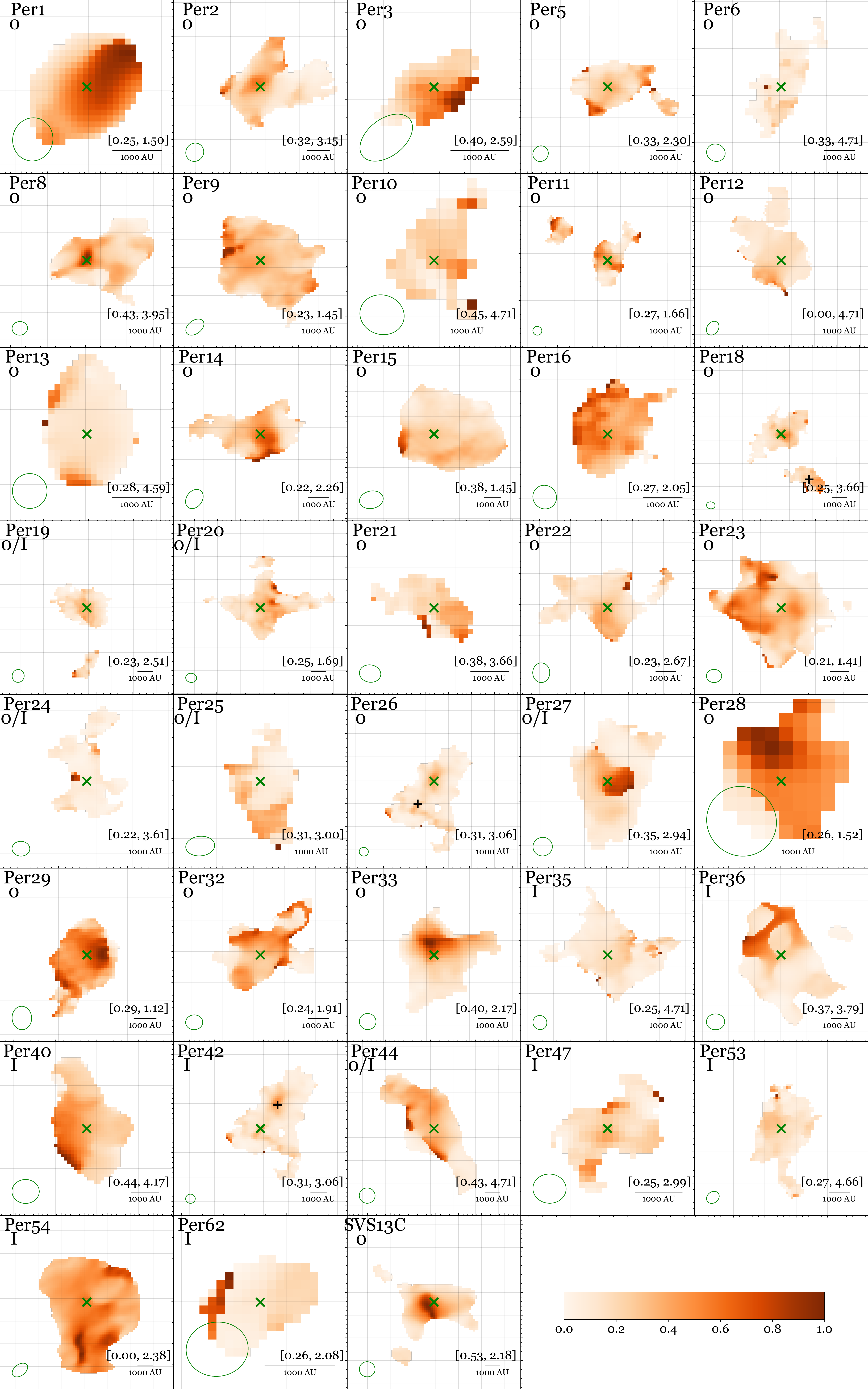}
    \caption{Full width at half maximum (FWHM) line widths of MASSES survey sources studied in this paper. The minimum (light orange) and maximum (dark orange) line width values are given in brackets in km s$^{-1}$, with a normalized linear colorbar shown in the bottom right. The synthesized beam is in the lower left in green. The center of the 2D Gaussian ellipse fit (see Figure \ref{fig:gallery_mom0}) is shown as a green X, with other sources in the map denoted as black +'s. All grids have 5" by 5" spacing, with a scalebar showing 1000 au in the lower right corner of each plot.}
    \label{fig:gallery_line width}
\end{figure*}

In this appendix we present the galleries of the moment 0, the velocity, and line width maps in Figures~\ref{fig:gallery_mom0}, \ref{fig:gallery_mom1}, \ref{fig:gallery_line width}, respectively.
%\section{Moment Maps} \label{appendix:A}

% Uncomment this when finished
%\begin{comment}

%\end{comment}

%% References

\bibliography{main}{}
\bibliographystyle{aasjournal}

\listofchanges

\end{document}